\documentclass[twocolumn, secnumarabic, amssymb, nobibnotes, aps, prd, superscriptaddress, nofootinbib]{revtex4-2}

\usepackage[left=2cm, right=2cm, top=2.5cm, bottom=2.5cm]{geometry}

\usepackage{amsfonts}
\usepackage{amssymb}
\usepackage{amsmath}
\usepackage{dcolumn}
\usepackage{physics}
\usepackage{appendix}
\usepackage{theorem}
\usepackage{lipsum}
\usepackage{mathtools}
\usepackage{float}
\usepackage{subcaption}

\usepackage{tabularx}
\usepackage{multirow}

\usepackage{hyperref}
\hypersetup{
    colorlinks=true,
    linkcolor=blue,     % eq and fig referencing
    citecolor=blue,     % ref citing
    urlcolor=blue,      % color of urls
}

\usepackage[version=4]{mhchem}

\let\oldAA\AA
\renewcommand{\AA}{\text{\normalfont\oldAA}}

\usepackage{xcolor}

\definecolor{mygreen}{rgb}{0.0, 0.5, 0.0}
\definecolor{myred}{rgb}{0.9, 0.05, 0.15}
\definecolor{nbblue}{rgb}{0.075, 0.0, 0.7}

\usepackage{natbib}
\begin{document}

%%%%%%%%%%%%%%%%%%%%%%%%%%%%%%%%%%%%%%%%%%%%%%%%%%%%%%%%%%%%%%%%%%%%%%%%%%%%%%%%

\title{On the Simulation of Conical Intersections in Water and Methanimine Molecules Via Variational Quantum Algorithms}

\author{Samir Belaloui}
\email{samir.belaloui@doc.umc.edu.dz}
\affiliation{Laboratoire de Physique Math\'{e}matique et Subatomique, Fr\`{e}res Mentouri University Constantine 1, Ain El Bey Road, Constantine, 25017, Algeria}%

\author{Nacer Eddine Belaloui}
\affiliation{Constantine Quantum Technologies, \\Fr\`{e}res Mentouri University Constantine 1, Ain El Bey Road, Constantine, 25017, Algeria}
\affiliation{Laboratoire de Physique Math\'{e}matique et Subatomique, Fr\`{e}res Mentouri University Constantine 1, Ain El Bey Road, Constantine, 25017, Algeria}%

\author{Achour Benslama}
\affiliation{Constantine Quantum Technologies, \\Fr\`{e}res Mentouri University Constantine 1, Ain El Bey Road, Constantine, 25017, Algeria}
\affiliation{Laboratoire de Physique Math\'{e}matique et Subatomique, Fr\`{e}res Mentouri University Constantine 1, Ain El Bey Road, Constantine, 25017, Algeria}%

%%%%%%%%%%%%%%%%%%%%%%%%%%%%%%%%%%%%%%%%%%%%%%%%%%%%%%%%%%%%%%%%%%%%%%%%%%%%%%%%%%%%%%%%%%%%%%%%%%%%%%%%
\begin{abstract}
    We investigate the electronic structure of methanimine (\ce{CH2NH}) and water (\ce{H2O}) molecules in an effort to locate conical intersections (CIs) using variational quantum algorithms. Our approach implements and compares a range of hybrid quantum-classical methods, including the Variational Quantum Eigensolver (VQE), Variational Quantum Deflation (VQD), VQE with Automatically-Adjusted Constraints (VQE-AC), and we explore molecular configurations of interest using a State-Average (SA) approach. Exact Diagonalization is employed as the classical benchmark to evaluate the accuracy of the quantum algorithms. We perform simulations across a range of molecular geometries, basis sets, and active spaces to compare each algorithm's performance and accuracy, and to enhance the detectability of CIs. This work confirms the quantum variational algorithms' capability of describing conical intersections in both molecules, as long as appropriate active spaces and geometries of the molecule are chosen. We also compare the accuracy and reliability of VQE-based methods for computing excited states with classical benchmark methods, and we demonstrate good agreement within desired regions.
\end{abstract}

\maketitle

%%%%%%%%%%%%%%%%%%%%%%%%%%%%%%%%%%%%%%%%%%%%%%%%%%%%%%%%%%%%%%%%%%%%%%%%%%%%%%%%%%%%%%%%%%%%%%%%%%%%%%%%
\section{Introduction}
Quantum simulation has emerged as a powerful computational paradigm that leverages the principles of quantum mechanics to model complex quantum systems. By exploiting quantum phenomena such as superposition and entanglement, quantum simulators offer the potential to solve problems that are computationally intractable for classical computers \cite{Nielsen2010,Feynman1982,Benioff1980,Manin1980}. A particularly promising area for quantum simulation is quantum chemistry, where the accurate computation of molecular electronic structures and energy spectra remains challenging \cite{Kassal2011}, especially for large molecules.
The accurate simulation of molecular systems--involving multiple electrons and molecular orbitals--presents a major computational hurdle due to the exponential scaling of the Hilbert space with system size \cite{Cai2020}. Classical methods attempt to overcome this challenge through various approximations. Among them, the Hartree-Fock method, for instance, treats each electron as moving independently in an average potential generated by the other electrons \cite{Hartree1928,Fock1930}. Density Functional Theory (DFT) reformulates the problem in terms of electron density rather than the many-electron wave function in terms of a function of only three spatial coordinates offering an efficient alternative for large systems \cite{Kohn1999}. Meanwhile, the Density Matrix Renormalization Group (DMRG) approach leverages a matrix product state representation to capture strong correlation effects in low-dimensional systems \cite{White1992,White1993}.
The most accurate results are obtained with the Full Configuration Interaction (FCI) method, a linear variational approach that fully accounts for electron correlation and, within a given basis set, can recover nearly all of the exact non-relativistic correlated energy for small systems \cite{Shavitt1981}.
Quantum computing offers a fundamentally new approach for tackling these problems. By encoding the molecular Hamiltonian into qubits, it becomes possible to simulate the quantum evolution of the system and extract eigenvalues corresponding to ground- and excited-state energies \cite{McArdle2020}. Determining these energy levels is essential for understanding chemical reactivity, stability, and spectroscopy \cite{neese2017}. 
Accurate molecular energy calculations are essential to a variety of applications, including the design of novel pharmaceuticals and catalysts, the development of advanced materials for energy storage and superconducting technologies, and the elucidation of chemical and biological reaction mechanisms \cite{Zini2023,Gross2012}. In recent years, several quantum algorithms have been developed to address the ground-state energy problem. Among them, the Variational Quantum Eigensolver (VQE) \cite{Peruzzo2014} stands out as a promising hybrid quantum-classical method, suited for near-term quantum devices. It uses the variational principle and Rayleigh--Ritz theorem \cite{Tilly2022,Cohen-Tannoudji2017} to iteratively optimize a parameterized quantum circuit. In contrast, Quantum Phase Estimation (QPE) \cite{Kitaev1995} provides high-precision results but requires beyond near term quantum hardware. Other notable approaches include Quantum Imaginary Time Evolution (QITE) \cite{Motta2020}, which simulates non-unitary dynamics to drive the system to its ground state, and Adiabatic Quantum Computing (AQC) \cite{Farhi2000}, which relies on the adiabatic evolution of a quantum system's Hamiltonian. These are some of the fundamental single-state quantum energy estimation methods and a starting point for more advanced excited-state extensions. This work focuses on computing the ground-state energy of small molecules using the VQE algorithm.\\
The computation of excited-state energies is equally important and more challenging. Classical techniques such as Configuration Interaction \cite{Roothaan1951}, Equation-of-Motion Coupled Cluster (EOM-CC) \cite{Stanton1993}, Time-Dependent Density Functional Theory (TD-DFT) \cite{Runge1984}, and multireference methods like CASSCF \cite{Roos1980} and MRCI \cite{Werner1988} are well-established. Many of these methods rely on state-averaging strategies to maintain accuracy across multiple states. On the quantum side, a variety of methods have been developed to extend ground-state algorithms to excited states. These include Variational Quantum Deflation (VQD) \cite{Grimsley2019}, Quantum Subspace Expansion (QSE) \cite{McClean2017}, State-Averaged VQE (SA-VQE) \cite{Yalouz2021}, and Quantum Equation-of-Motion (qEOM) \cite{Ollitrault2020}. Other strategies, such as the Folded Spectrum method \cite{cadi2024} and VQE with Adjusted Constraints (VQE-AC) \cite{Gocho2023}, offer alternative routes to accessing higher-energy eigenstates. While QPE also provides access to excited states, its hardware demands limit its current applicability \cite{Kitaev1995}. Together, these methods form a diverse toolkit for exploring excited-state phenomena across classical and quantum computing platforms.\\
In molecular systems, a conical intersection arises when two adiabatic potential energy surfaces associated with different electronic states become degenerate and intersect in a multidimensional nuclear coordinate space \cite{Gonon2017,Yarkony1996}. The resulting topography near the intersection resembles a pair of inverted cones sharing a common apex. This degeneracy facilitates ultrafast nonradiative transitions between electronic states that are crucial in a diverse array of chemical, physical, and biological processes initiated by electronic excitation. Consequently, the accurate theoretical treatment of conical intersections constitutes a significant and ongoing area of research and development within the field. Ultimately, the understanding and controlled manipulation of molecular behavior through conical intersections are key objectives in emerging fields such as photocatalysis, organic electronics, and photomedicine \cite{Levine2007,Rivera2021,Zhou2023}.
In this work, we explore the capabilities of variational quantum algorithms in detecting conical intersections in small molecular systems. Our aim is to identify the algorithmic components and strategies essential for reliably characterizing these nonadiabatic features.\\\\
This paper is organized as follows: In Section \ref{seq:theoretical-background}, we define the general form of the molecular Hamiltonian, introducing concepts such as second quantization and the various mappings required to represent fermionic systems on qubits. This section also contains an overview of the concepts of conical intersections, avoided crossings, and the relation with the Born-Oppenheimer approximation. In Section \ref{seq:algorithms},  we provide an overview of several variational quantum algorithms, specifically discussing the Variational Quantum Eigensolver (VQE), Variational Quantum Deflation (VQD), VQE with automatically-adjusted constraints  (VQE-AC). In Section \ref{seq:computational-details} we detail the computational specifics of our study, encompassing the choice of basis sets, the design of the quantum ansatz, and the selection of the classical optimizer. Then we describe each of the 4 simulation sets performed in this study. Section \ref{seq:results} is dedicated to the presentation of our results and a discussion of their implications and significance. Finally, the paper's conclusion is presented in Section \ref{seq:conclusion}.

%%%%%%%%%%%%%%%%%%%%%%%%%%%%%%%%%%%%%%%%%%%%%%%%%%%%%%%%%%%%%%%%%%%%%%%%%%%%%%%%%%%%%%%%%%%%%%%%%%%%%%%%
\section{Theoretical Background}
\label{seq:theoretical-background}
%%    %%    %%    %%    %%    %%    %%    %%    %%    %%    %%    %%    %%    %%    %%    %%    %%    %%
\subsection{The Molecular and Electronic Hamiltonians}
The general molecular Hamiltonian that describes the dynamics of a molecule in the system of atomic units is \cite{Szabo1996}
\begin{multline}
    H_\text{mol} = - \sum_n \frac{\Delta_n}{2M_n} - \sum_i \frac{\Delta_i}{2} - \sum_{i,m} \frac{Z_m}{|\mathbf{R}_m - \mathbf{r}_i|}\\
    + \sum_{m,n>m} \frac{Z_m Z_n}{|\mathbf{R}_n - \mathbf{R}_m|} + \sum_{i,j>i} \frac{1}{|\mathbf{r}_i - \mathbf{r}_j|},
\end{multline}
where $m,n$ and $i,j$ are used to sum over nuclei and electrons respectively, $\mathbf{R}_i$ are nuclear positions, $\mathbf{r}_i$ are electronic positions, and $M_i$ are nuclear masses.\\
Due to the large difference in nuclear and electronic masses, we can use the Born--Oppenheimer approximation, which consists of neglecting the parts of the molecular Hamiltonian corresponding to the motion of nuclei. We obtain the  simpler electronic Hamiltonian:
\begin{equation}
    H_\text{elect} = - \sum_i \frac{\Delta_i}{2} - \sum_{i,m} \frac{Z_m}{|\mathbf{R}_m - \mathbf{r}_i|} + \sum_{i,j>i} \frac{1}{|\mathbf{r}_i - \mathbf{r}_j|},
    \label{eq:elect-ham}
\end{equation}
where the $\sum_{m,n>m} \frac{Z_m Z_n}{|\mathbf{R}_n - \mathbf{R}_m|}$ term became a nuclear repulsion energy constant.\\

\noindent The electronic Hamiltonian \eqref{eq:elect-ham} can be expressed in terms of the creation and annihilation operators $a^\dagger$, $a$ as \cite{Szabo1996}
\begin{equation}
    H_{elect}^{2^{\rm nd}}=\sum_{p,q}^{}h_{pq}a_{p}^{\dagger}a_{q}+\sum_{p,q,r,s}^{}h_{pqrs}a_{p}^{\dagger}a_{q}^{\dagger}a_{r}a_{s}
\end{equation}
where $a_p^\dagger$ and $a_p$ create and annihilate electrons in the $p^\text{th}$ orbital. Whereas we define
\begin{equation}
    h_{pq}=\int d\mathbf{x} \phi_{p}^{*}(\mathbf{x})\left( -\frac{\Delta }{2}-\sum_{i}\frac{Z_{i}}{| \mathbf{R}_{i}-\mathbf{r} |} \right) \phi_{q}(\mathbf{x})
\end{equation}
and
\begin{equation}
    h_{pqrs}=\int d\mathbf{x}_{1}d\mathbf{x}_{2}\frac{\phi_{p}^{*}( \mathbf{x}_{1} )\phi_{q}^{*}( \mathbf{x}_{2} )\phi_{r}( \mathbf{x}_{1} )\phi_{s}( \mathbf{x}_{2} )}{| \mathbf{r}_{1}-\mathbf{r}_{2} |}
\end{equation}
as the one- and two-electron integrals, respectively. Lastly $\mathbf{x}_i = (\mathbf{r}_i, s_i)$ describes both the position and spin of an electron, and $\phi_i(\mathbf{x})$ is a molecular spin orbital wave function.\\

In this paper, to construct the $\phi(\mathbf{x})$ wave functions, we employed the STO-3G and 6-31G basis sets, which are mathematical functions used to approximate molecular wave functions via the Linear Combination of Atomic Orbitals (LCAO) method \cite{Levine1999}. The full second quantization procedure was performed via the python Qiskit \cite{qiskit} package, which in turn calls the Python for Strongly Correlated Electron Systems (PySCF) package \cite{Sun2020}.

%%    %%    %%    %%    %%    %%    %%    %%    %%    %%    %%    %%    %%    %%    %%    %%    %%    %%
\subsection{The Conical Intersection and Avoided Crossings}
Conical intersections represent fundamental concepts within quantum chemistry, serving as the mechanistic foundation for understanding ultrafast non-radiative transitions \cite{Gonon2017,Yarkony1996}. Indeed, when a molecule undergoes an electronic excitation following absorbed radiation for example, it evolves in an excited state and ends up returning to its ground state by radiation emission. At a conical intersection however, this dissipation of energy is no longer done by radiation but by its transfer in the form of vibronic energy of the molecule \cite{Worth2004}.\\
The presence of this conical intersection places us far from the domain of validity of the Born--Oppenheimer (BO) approximation. In fact, the Schrödinger equation corresponding to a molecular system results in a set of coupled equations; the Born–Oppenheimer approximation consists of neglecting the coupling terms and thus allows the separation of nuclear and electronic motions. This is justified by the fact that the coupling terms are inversely proportional to the nuclear masses, which are significantly larger than that of the electrons \cite{BornOppenheimer1927, Gonon2017}. The problem is that these coupling terms are at the same time proportional to factors of the form \cite{Gonon2017,Yarkony1996}
$$\frac{1}{\epsilon_{i}(R)-\epsilon_{j}(R)},$$
and in particular, in the case of the ground and first excited states, this factor is
$$\frac{1}{\epsilon_{0}(R)-\epsilon_{1}(R)},$$
where $R$ represents the set of coordinates of the molecule's nuclei, which essentially describes its geometry, and $\epsilon_{i}(R)$ are the eigenvalues associated with the electronic states\\ 
It is clear that the closer two electronic states are, the more this term increases, until it diverges at the point of energy degeneracy. This leads us to consider a few observations regarding the validity of the Born--Oppenheimer approximation in the vicinity of this point. In molecular systems, avoided crossings arise when two potential energy surfaces approach each other closely but do not intersect, typically as a result of the Born–Oppenheimer approximation, which treats electronic and nuclear motions separately. Within this framework, electronic states are considered adiabatic and represented as eigenfunctions of the electronic Hamiltonian for fixed nuclear positions. Consequently, when two electronic states come energetically close, they repel due to the non-crossing rule, resulting in an avoided crossing rather than a true intersection \cite{vonNeumann1929, Baer2006}. This phenomenon reflects the limitations of the BO approximation, particularly in regions of strong nonadiabatic coupling \cite{Yarkony1996}. In polyatomic systems, an avoided crossing is often interpreted as evidence for the presence of an underlying conical intersection \cite{Domcke2004}. However, in practical quantum chemical calculations, the ability to locate conical intersections depends critically on the method and level of theory employed. It is often the case that, due to an insufficient active space or limitations in the electronic structure method, the conical intersection cannot be explicitly captured, and only an avoided crossing is revealed \cite{Gonzalez2012}. This limitation challenges the accurate modeling of nonadiabatic dynamics governed by conical intersections \cite{Yarkony2001}.

%%    %%    %%    %%    %%    %%    %%    %%    %%    %%    %%    %%    %%    %%    %%    %%    %%    %%
\subsection{The Mapping}
Qubit mapping is a technique used to map fermionic operators onto qubit operators for applications such as quantum chemistry simulations. This allows us to represent fermionic systems, which describe particles like electrons, for use on quantum computers and to operate on qubits. In this technique, we represent the creation and annihilation operators $a^{\dagger}_{i}$, $a_{i}$, in the second-quantized Hamiltonian as Pauli strings $\left\{ I,X,Y,Z \right\}$. In other words, qubit mapping establishes a correspondence between electrons in spin-orbitals and qubits.
Several methods can be employed to achieve this mapping. These include parity mapping, Jordan-Wigner mapping, and Bravyi-Kitaev mapping \cite{Tilly2022}, which are among the most popular. In this paper, we have used the parity mapping.\\
In parity mapping, we encode on each qubit, $i$, the parity of electron occupation, specifically representing whether an even or odd number of electrons occupy the spin orbitals from the first orbital up to the $i$-th orbital. \cite{Bravyi2002,Seeley2012}.
One of the primary advantages of parity mapping is its ability to reduce the qubit requirements by 2, and thus minimizing the number of gates needed for quantum circuit operations.
The transformations for parity mapping are given by \cite{Tilly2022}
\begin{equation}
    a_{j}\to \frac{Z_{j-1}\bigotimes X_{j}+i Y_{j}}{2}\bigotimes X_{j+1}\bigotimes \dots \bigotimes X_{n-1}
\end{equation}
\begin{equation}
    a_{j}^{\dagger}\to \frac{Z_{j-1}\bigotimes^{}X_{j}-i Y_{j}}{2}\bigotimes_{}^{} X_{j+1}\bigotimes_{}^{}...\bigotimes_{}^{}X_{n-1}
\end{equation}
where $n$ is the number of qubits corresponding to the number of spin orbitals.

%%%%%%%%%%%%%%%%%%%%%%%%%%%%%%%%%%%%%%%%%%%%%%%%%%%%%%%%%%%%%%%%%%%%%%%%%%%%%%%%%%%%%%%%%%%%%%%%%%%%%%%%
\section{Algorithms and Methods}
\label{seq:algorithms}
%%    %%    %%    %%    %%    %%    %%    %%    %%    %%    %%    %%    %%    %%    %%    %%    %%    %%
\subsection{Overview of the Variational Quantum Eigensolver}
\begin{figure}[t]
    \centering
    \includegraphics[width=0.95\linewidth]{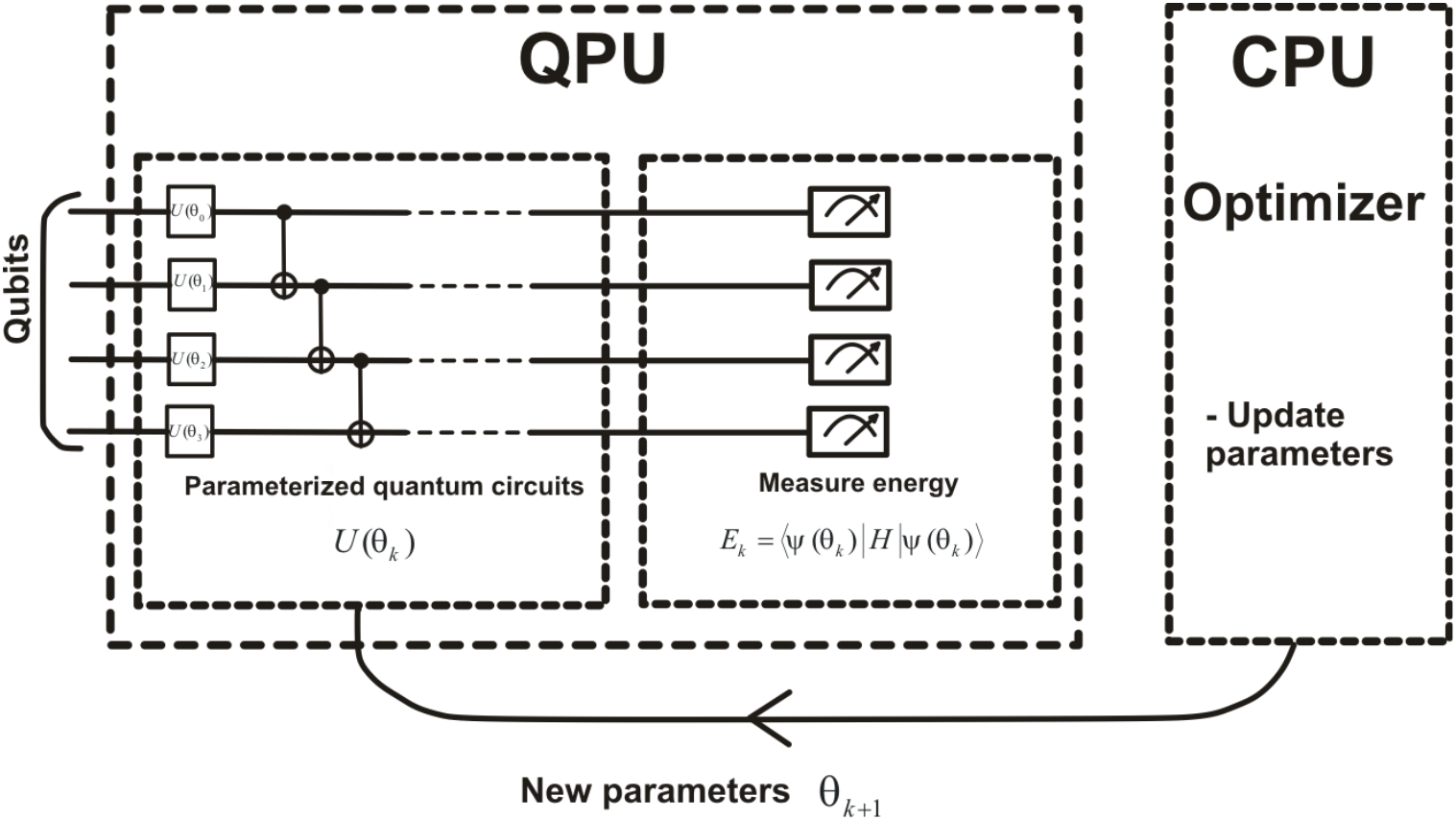}
    \caption{
        VQE algorithm:  the Quantum Processing Unit (QPU) runs a parameterized quantum circuit to prepare a state and measure its energy. This energy value is then sent to a Classical Processing Unit (CPU) that optimizes the circuit parameters. The updated parameters are fed back to the QPU in an iterative process, aiming to minimize the energy. Figure adapted from Belaloui et al.~\cite{Belaloui2025}.
}
    \label{fig:vqe-loop}
\end{figure}

The Variational Quantum Eigensolver (VQE) is a hybrid quantum-classical algorithm designed to estimate the lowest possible energy of a quantum system, known as the ground-state energy. It is based on the Rayleigh-Ritz theorem in quantum mechanics, which states that, for any normalized trial wavefunction $\ket{\Psi\left( \theta \right)}$, the calculated energy will always be equal to or higher than the true ground state energy $E_{0}$ \cite{Tilly2022,Cohen-Tannoudji2017}. More explicitly,
\begin{equation}
    E\left( \theta \right)=\bra{\Psi\left( \theta \right)}H\ket{\Psi\left( \theta \right)}\ge E_{0}.
\end{equation}
Within the VQE framework, the term "ansatz" usually refers to the quantum circuit used to construct the trial wave function. It is a template with variable components that one can modify in an attempt to approximate the actual ground state of a quantum system.
In practice, the VQE uses a quantum computer to evaluate the energy observable of an ansatz' parameterized quantum circuit. This quantum state is intended to converge to the true ground state of the system. The quantum computer is thus used to measure the ansatz corresponding energy, and the measurement outcomes are in turn used to fine-tune the parameters of the circuit by a classical computer.
As described in Fig. \ref{fig:vqe-loop}, this process is iteratively repeated: the quantum computer prepares the quantum state and performs the energy measurements, and classical computer adjusts the parameters to reduce the estimate of the energy. Going through a sufficient number of iterations, the process attempts to converge to the optimal set of parameters producing the lowest possible estimate of the energy, hopefully equal to the ground-state energy of the real system.\\
The VQE is particularly of interest for near-term quantum computers as it keeps quantum computations relatively short and relies on classical computation to handle the more complex part of the optimization.

We may summarize the main steps of the VQE as follows:

\begin{itemize}
    \item Hamiltonian Construction: The molecular Hamiltonian is constructed using quantum chemistry methods, such as Hartree--Fock or density functional theory. This involves representing the electronic interactions and nuclear potential energy.
    \item Qubit Encoding: The Hamiltonian is mapped onto the qubits. 
    \item Ansatz Circuit Design: The ansatz is a parametrized quantum circuit, which is designed to prepare a trial state. This ansatz circuit takes in a set of parameters, often denoted as $\theta$, and outputs a quantum state $\ket{\Psi\left( \theta \right)}$.
    \item Expectation Value Measurement: We measure $\bra{\Psi\left( \theta \right)}H\ket{\Psi\left( \theta \right)}$ which is the energy of the trial state $\ket{\Psi\left( \theta \right)}$.
    \item Classical Optimization: The measured expectation value $\bra{\Psi\left( \theta \right)}H\ket{\Psi\left( \theta \right)}$ is then used by a classical optimizer to adjust the parameters $\theta$ of the ansatz circuit, aiming to minimize the cost function, which is, in our case, the energy. The optimizer iteratively refines the parameters $\theta$ and feeds them back to the quantum computer to generate a new trial quantum state $\ket{\Psi\left( \theta \right)}$.
    \item Convergence Check: The process continues until the energy stabilizes or a predefined stopping threshold is reached. The final minimum energy obtained is, in general, an approximation of the ground-state energy of the system.
\end{itemize}

%%    %%    %%    %%    %%    %%    %%    %%    %%    %%    %%    %%    %%    %%    %%    %%    %%    %%
\subsection{Overview of Variational Quantum Deflation Algorithm}

The VQE is used primarily to determine the ground state of a quantum system. Nevertheless, numerous applications, particularly in quantum chemistry and materials science, necessitate understanding the system's excited states, which represent higher energy levels.
Therefore, to access these excited states, we turn to the Variational Quantum Deflation (VQD) algorithm. The VQD algorithm determines excited states by iteratively modifying the variational ansatz to eliminate contributions from previously obtained lower-energy states. This is achieved by introducing penalty terms into the cost function that suppress overlap with known eigenstates, thereby guiding the optimization toward higher-energy solutions. This penalty is designed to be large when the current variational state has a significant overlap with the previously found ground state. A common form of the cost function is \cite{Higgott2019}: 
\begin{equation}
    F\left( \theta_{k} \right)=\bra{\Psi\left( \theta_{k} \right)}H\ket{\Psi\left( \theta_{k} \right)}+ \sum_{i=1}^{k-1} \beta_{i} \left| \bra{\Psi\left( \theta_{k} \right)} \ket{\Psi\left( \lambda_{i} \right)} \right|^{2},
    \label{eq:vqd-cost}
\end{equation}
where $F\left( \theta_{k} \right)$ denotes the cost function to be minimized. The parameters $\theta_{k}$ represent the variational parameters of the quantum circuit (ansatz) used to prepare the $k^{\text{th}}$ excited state, denoted by $\ket{\Psi\left( \theta_{k} \right)}$. The goal of the VQD algorithm is to find the values of $\theta_{k}$ that minimize the cost function $F\left( \theta_{k} \right)$. The expression $\bra{\Psi\left( \theta_{k} \right)} H \ket{\Psi\left( \theta_{k} \right)}$ is the expectation value of the Hamiltonian $H$ with respect to the quantum state $\ket{\Psi\left( \theta_{k} \right)}$. The term $\sum_{i=1}^{k-1} \beta_{i} \left| \bra{\Psi\left( \theta_{k} \right)} \ket{\Psi\left( \lambda_{i} \right)} \right|^{2}$ is the penalty term that enforces orthogonality with the previously found $k$ states (from the ground state, $i=0$, up to the $(k-1)^{\text{th}}$ excited state).
In practice we choose sufficiently large values of $\beta$ such that the penalty term becomes sufficiently significant in the cost function \eqref{eq:vqd-cost}, but not so large that it makes the energy term irrelevant in the optimization.

%%    %%    %%    %%    %%    %%    %%    %%    %%    %%    %%    %%    %%    %%    %%    %%    %%    %%
\subsection{Overview of The Variational Quantum Eigensolver with Automatically-Adjusted Constraints (VQE-AC)}

The Variational Quantum Eigensolver with automatically-adjusted constraints (VQE-AC) \cite{Gocho2023} is an extension of the traditional VQE algorithm that incorporates adjustable constraints into the optimization process. In many computational chemistry or material science problems, adding constraints helps ensure that the solution meets specific physical or chemical requirements, such as particle number conservation, spin constraints, or symmetry considerations. 
Unlike the VQD's cost function, this method's cost function is simply the Hamiltonian’s expectation value
\begin{equation}
    E\left( \theta \right)= \bra{\Psi\left( \theta \right)}H \ket{\Psi\left( \theta \right)}
    \label{eq:vqe-ac-cost}
\end{equation}
subject to an externally-handled orthogonality constraint
\begin{equation}
    \left| \bra{\Psi\left( \theta \right)} \ket{\Psi_{0}}\right|^{2}\le 10^{-4}.
    \label{eq:vqe-ac-constraint}
\end{equation}
The VQE-AC employs optimizers such as COBYLA \cite{Powell1994} that explicitly impose this constraint between the ground and excited states, without requiring manual tuning of a penalty parameter as in the VQD. The constraint approach enhances stability and accuracy of the optimization, especially in the vicinity of conical intersections, since tiny parameter changes can result in state mixing or erroneous state assignment. In the VQD, improper setting of the penalty weight can cause the optimizer to converge to a wrong state or lose orthogonality, especially in such sensitive regions \cite{Gocho2023}.

%%    %%    %%    %%    %%    %%    %%    %%    %%    %%    %%    %%    %%    %%    %%    %%    %%    %%
\subsection*{On The State-Average Method}
The State-Average Method (SA) is a classical technique used in quantum chemistry to compute properties of multiple quantum states simultaneously, such as in excited-state calculations. It is especially useful when one wants to study not just the ground state of a system, but also excited states, or when several quantum states are relevant to a problem \cite{Yalouz2021}. We use the SA method as a means to explore molecular configurations of interest in the search for conical intersections.

%%%%%%%%%%%%%%%%%%%%%%%%%%%%%%%%%%%%%%%%%%%%%%%%%%%%%%%%%%%%%%%%%%%%%%%%%%%%%%%%%%%%%%%%%%%%%%%%%%%%%%%%
\section{Computational Details}
\label{seq:computational-details}
%%    %%    %%    %%    %%    %%    %%    %%    %%    %%    %%    %%    %%    %%    %%    %%    %%    %%
\subsection{Basis Sets}
Exactly solving the Schrödinger equation for molecules in quantum chemistry is a typically intractable problem. To circumvent this, calculations are simplified by representing atomic orbitals--the solutions for single atoms--as combinations of more manageable functions, which are collectively known as basis sets. In this paper, we employ two distinct types of basis sets to approximate these atomic orbitals.\\
The first type is the STO-3G (Slater-Type Orbital - 3 Gaussian) basis \cite{lewars2010-basis-sets}, where each atomic orbital is approximated by a fixed linear combination of three Gaussian-type functions. These Gaussian functions are combined to mimic the behavior of a Slater-Type Orbital (STO), which more closely resembles the true shape of an atomic orbital.\\
The second basis set utilized is the 6-31G \cite{lewars2010-basis-sets}, a type of split-valence basis set. With 6-31G, core atomic orbitals are represented by a contraction of six Gaussian-type functions. In contrast, valence atomic orbitals, which are crucial for chemical bonding, are represented by two separate sets of Gaussian-type functions, each possessing different exponents. This splitting allows for a more flexible and accurate description of the valence electron distribution, reflecting the varying environments they might experience in a molecule.\\
It is worth noting that the number of qubits employed later in our quantum circuits directly corresponds to the count of spin orbitals explicitly represented within the qubit Hamiltonian. Given that this Hamiltonian is derived from a chosen basis set of atomic orbitals, the number of qubits required by the quantum algorithm can be interpreted as the effective size of the active space within the quantum algorithms, limiting the complexity of the electronic states that can be encoded and explored.

%%    %%    %%    %%    %%    %%    %%    %%    %%    %%    %%    %%    %%    %%    %%    %%    %%    %%
\subsection{Ansatz}
The formulation of the ansatz is of the highest priority. If it is too simple, it will be inaccurate. If it is too complex, it will be hard to run on current quantum computing hardware and make the optimization more complex.\\
Two standard forms of ansatz are considered. The first, the hardware-efficient ansatz, is created to adapt to the constraints of contemporary quantum hardware and performs effectively but might not be extremely precise. The second form, the chemically-motivated ansatz, employs knowledge of the physics of the chemical problem to construct improved estimates, thereby enhancing precision at the cost of greater implementation complexity \cite{Belaloui2025}.\\
Selecting an appropriate ansatz involves a delicate trade-off between accuracy and viability. It must be robust enough to yield satisfactory outcomes, but also be simple enough to implement and optimize practically.\\
Explicitly, we will use:
\begin{itemize}
    \item The Efficient SU2 ansatz, a specific type of hardware-efficient ansatz, constructed through the Qiskit quantum computing framework. Efficient SU2 is commonly used to prepare trial states in VQE for quantum chemistry calculations \cite{Tilly2022,Belaloui2025}.
    \item The Unitary Coupled Cluster Singles and Doubles (UCCSD) ansatz is a quantum chemistry technique used to construct an approximate quantum state that describes the electronic structure of a molecule. It is derived from the known coupled cluster theory \cite{Watts1989}, but adapted to be applied on quantum computers by imposing unitarity to suit quantum operations \cite{Barkoutsos2018}.
    The SD refers to the electronic transitions the method is modeling: single excitations, where a single electron is excited, and double excitations, where two electrons are excited in a correlated manner. These are employed to describe the system's ground state in an incremental manner through successive excitations.
\end{itemize}

%%    %%    %%    %%    %%    %%    %%    %%    %%    %%    %%    %%    %%    %%    %%    %%    %%    %%
\subsection{Optimizers}
Classical optimization procedures are algorithms that direct the best parameter search in an experimental quantum setup. The algorithms check outcomes from quantum measurements and come up with alterations in the experimental configuration to achieve in the subsequent round, seeking to optimize the results, ultimately aiming towards discovering the minimum energy state or the optimal solution \cite{McClean2016}.
In VQAs, the main idea is that the quantum computer is tasked with preparing and measuring the states, while the classical optimizer decides how to update the settings for the next step based on those results. In this work we have used 3 different optimizers based on their performance and the requirements of the considered VQAs. Namely, these are SPSA, SLSQP, and COBYLA.\\\\
\noindent\textbf{SPSA (Simultaneous Perturbation Stochastic Approximation)} is a gradient-based optimizer \cite{Spall1992} well-suited for problems where analytical gradient calculation is challenging, particularly when encountering noisy data, a common characteristic in quantum computing. Unlike methods that compute the full gradient, SPSA efficiently approximates it using just two measurements per iteration of the cost function (in our case, the Hamiltonian expectation value), regardless of the number of variational parameters. This computational efficiency makes it a frequently employed technique in VQAs \cite{Pellow-Jarman2023}.

\noindent\textbf{SLSQP (Sequential Least Squares Programming)} in contrast is another gradient-based optimization algorithm \cite{Kraft1988}, but it is specifically tailored for constrained nonlinear optimization problems, making it suitable for applications like the VQD where penalty terms often introduce such constraints.

\noindent\textbf{COBYLA (Constrained Optimization By Linear Approximation)} is used in situations where derivative information is unavailable or unreliable. It offers a robust alternative. This numerical optimization method addresses problems by generating simple linear approximations of the objective and constraint functions, derived from a cluster of sampled points. These models are then used to compute variable changes that improve the solution. A "trust region" mechanism ensures that these changes remain within reasonable limits, maintaining the validity of the models. The algorithm iteratively updates its models and systematically progresses towards the optimum \cite{Powell1994}. COBYLA proves particularly useful for small to medium-sized problems where derivative information is either inaccessible or untrustworthy. Among the three optimizers we have used, COBYLA is essential for the VQE-AC algorithm \cite{Gocho2023}.

%%    %%    %%    %%    %%    %%    %%    %%    %%    %%    %%    %%    %%    %%    %%    %%    %%    %%
\subsection{Experiments Setup}
\label{subseq:exp-setup}
We carry a series of simulations on the \ce{H2O} and \ce{CH2NH} molecules, at different geometries, and with different active space sizes. We cluster the simulations into 4 distinct experiments focusing on several aspects of our work. The chosen molecular geometries and deformations are shown below, and the experimental configurations are explicitly given as well. We elected to first validate the quantum algorithms using the water molecule, then we moved on to investigating the behavior of avoided crossing points and conical intersections within specific geometrical deformations of the \ce{CH2NH} and \ce{H2O} molecules.\\

\textbf{Molecular geometries}\\

\begin{figure}[t]
    \centering

    \begin{subfigure}[b]{0.48\linewidth}
        \centering
        \includegraphics[width=\linewidth]{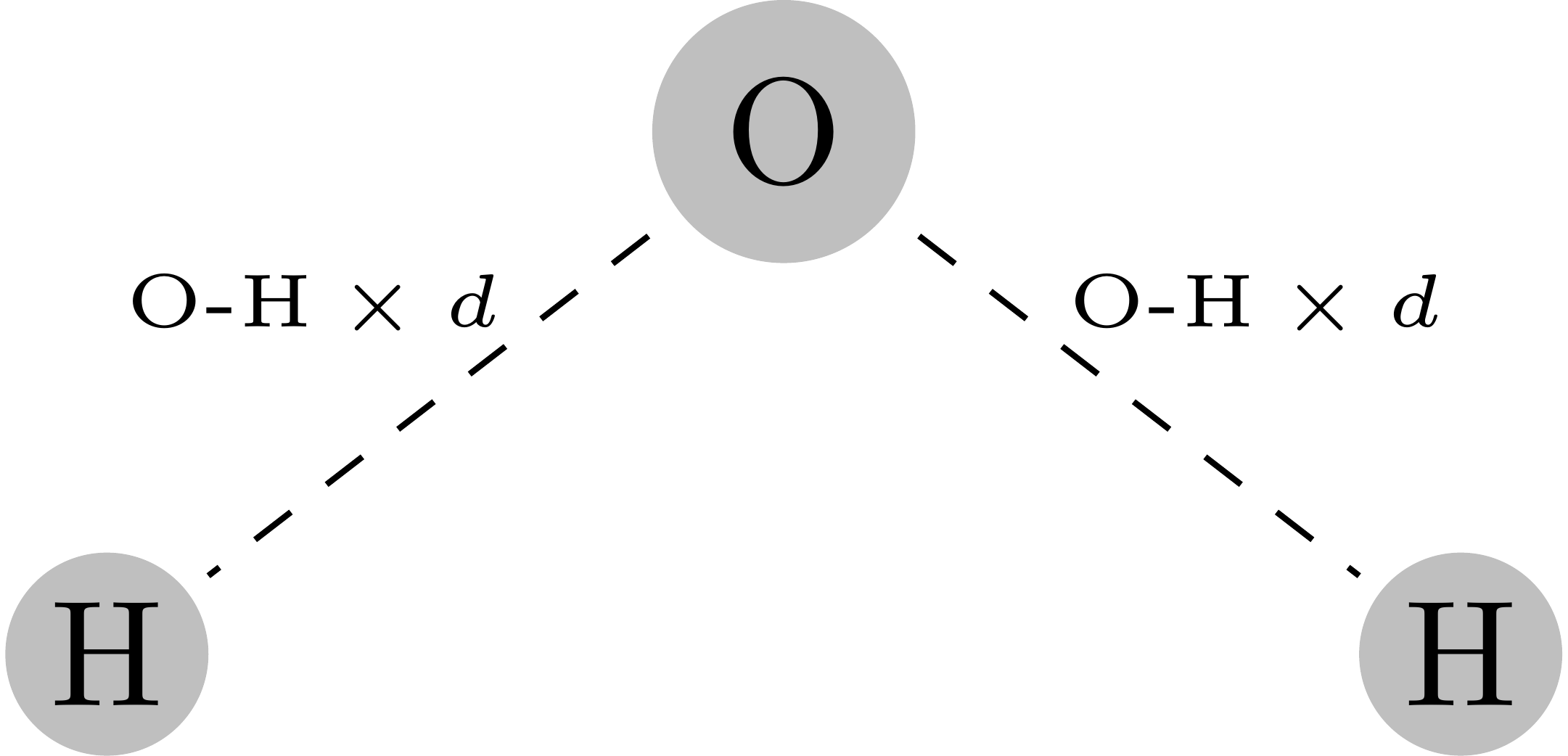}
        \caption{Deforming the \ce{H2O} molecule from its equilibrium geometry by scaling the \ce{O}--\ce{H} bond distances.}
        \label{fig:h2o-sym}
    \end{subfigure}
    \hfill
    \begin{subfigure}[b]{0.48\linewidth}
        \centering
        \includegraphics[width=\linewidth]{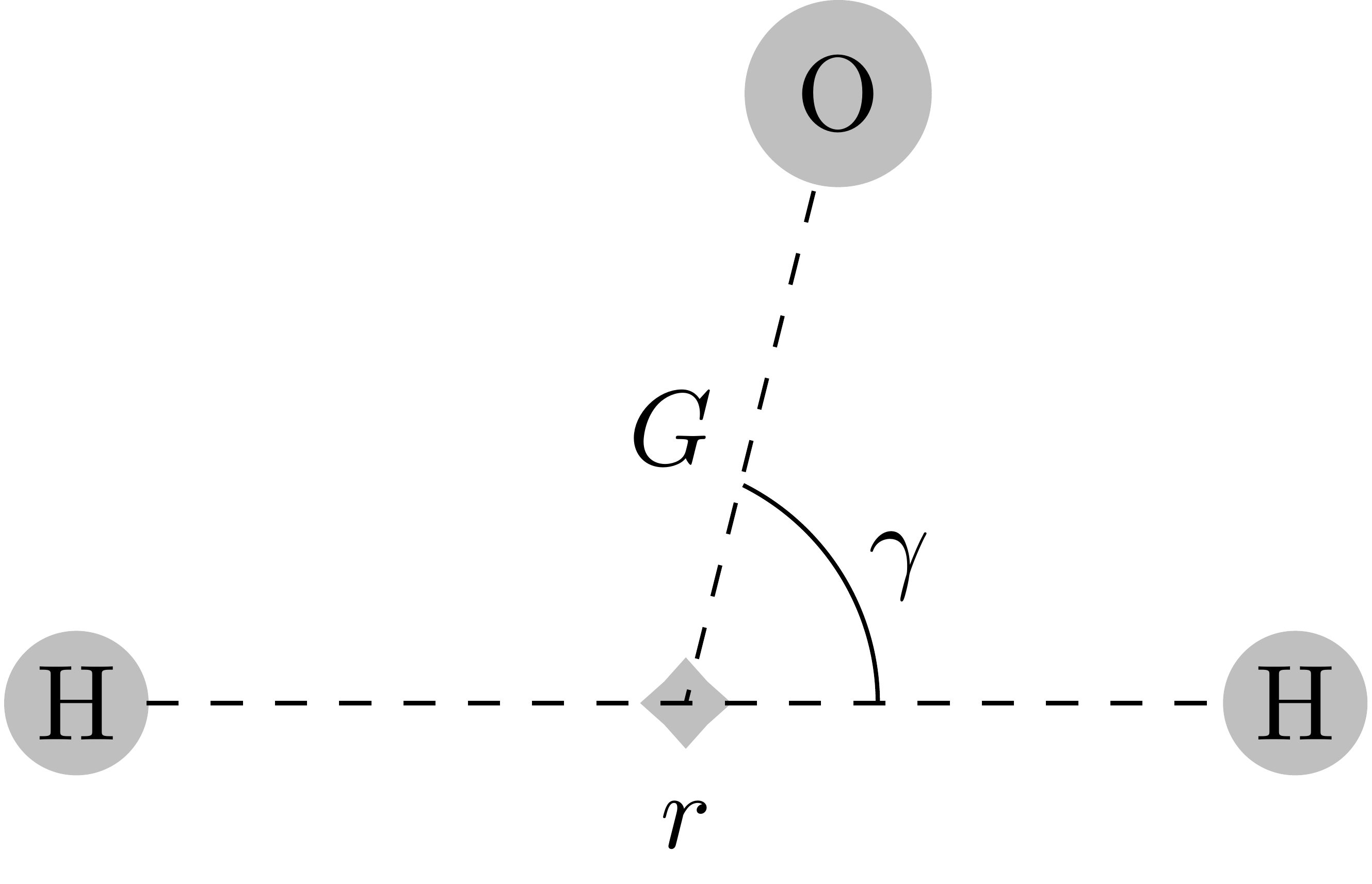}
        \caption{The water molecule in the Jacobi coordinate system. The geometry is defined around the \ce{H}--\ce{H} barycenter.}
        \label{fig:h2o-jacobbi}
    \end{subfigure}

    \caption{Geometrical representations of the \ce{H2O} molecule using two different deformation schemes.}
    \label{fig:h2o-combined}
\end{figure}

\begin{figure}[t]
    \centering
    \includegraphics[width=0.75\linewidth]{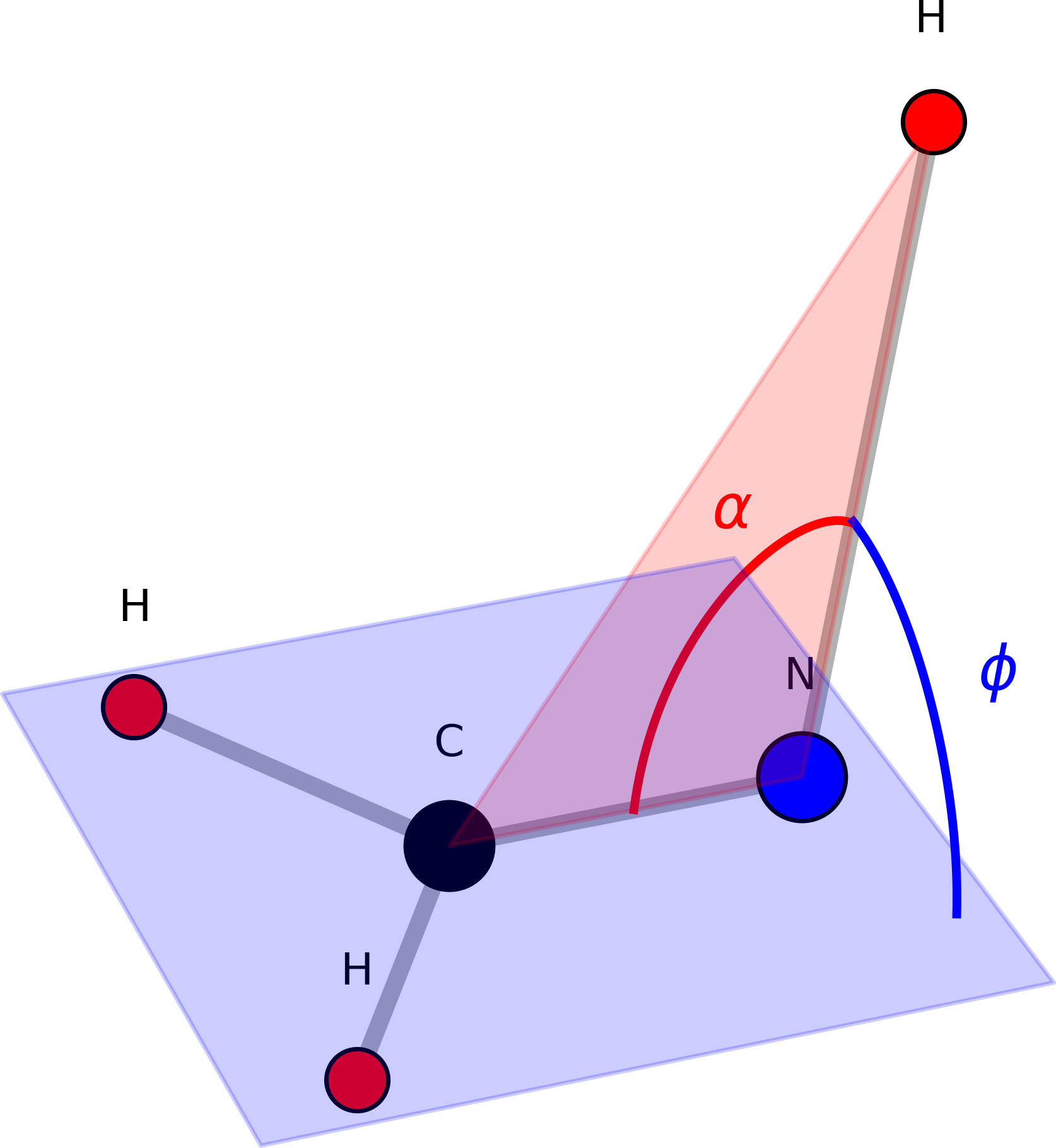}
    \caption{The used geometry of the Methanimine molecule with a varying bending angle $\alpha$ and a dihedral angle $\phi = 90^\circ$. The N, C, H, H atoms are coplanar (blue plane).}
    \label{fig:ch2nh-geometry}
\end{figure}

In this paper, we work with 3 geometries for the two aforementioned molecules. We use two coordinate systems and geometries for \ce{H2O}: a cartesian system where we scale up and down the O-H bond distances around the equilibrium geometry, as shown in Fig. \ref{fig:h2o-sym}, and a Jacobbi coordinate system with a heavier deformation, shown in Fig. \ref{fig:h2o-jacobbi}, that allows for the appearance of conical intersections in the water molecule’s spectrum. Jacobi coordinates are a set of coordinates employed to describe the relative motion of a system's constituent particles. For a water molecule, the Jacobi coordinates are defined as follows:
\begin{itemize}
     \item $r$: The distance between the two hydrogen atoms. It is fixed at r = 2.5832 $\AA$ \cite{Snyder2011}.
     \item $G$: The distance between the oxygen atom and the center of mass of the two hydrogen atoms.
     \item $\gamma$ The angle between the two hydrogen atoms and the oxygen atom. For this study, $\gamma$ is fixed at $0.00021^\circ$ \cite{Snyder2011}. 
\end{itemize}
For the methanimine molecule, the geometry is described in Fig. \ref{fig:ch2nh-geometry}. The important feature in this geometry is that the plane defined by the C-N-H atoms is perpendicular to the H-C-H plane, that is, the dihedral angle $\phi = 90^\circ$. We then vary the bending angle $\alpha$ in our simulations in the $[60^\circ, 160^\circ]$ interval.

\textbf{Experimental configurations}\\

\noindent\textbf{Notation:} In the following,  we will refer to active space configurations using the number of electrons and molecular orbitals as ($N_\textit{electrons}$, $N_\textit{orbitals}$)\\

We prepare the setup for each set of experiments as follows:
\begin{itemize}
    \item \textbf{Experiment 1 -- Validation of VQE, VQD, VQE-AC:} Starting from the equilibrium \ce{H2O} geometry, we scale the O-H bond distances by a scaling factor d. We use a 4-electron and 3-molecular (4, 3) orbital active space, as well as the STO-3G basis set and the Efficient SU2 ansatz for all algorithms. SPSA is chosen as the optimizer for the VQE and VQD, but COBYLA has to be used for the VQE-AC as it allows for the constraint adjustments. We also compare the VQE’s convergence in the absence and presence of simulated noise. For the VQD, the $\beta$ hyperparameter was set to 0.5.

    \item \textbf{Experiment 2 -- Conical intersection in the \ce{CH2NH} via VQD:} We investigate the use of the VQD for uncovering conical intersections in the \ce{CH2NH} system (see Fig. \ref{fig:ch2nh-geometry}). While using the Efficient SU2 ansatz, we prepare a (4, 3) then a (12, 9) active space and simulate the VQD in noiseless conditions using the SLSQP optimizer. We also set the $\beta$ hyperparameter in this case to $0.3$.

    \item \textbf{Experiment 3 -- Conical intersection in the \ce{CH2NH} via VQE-AC:} We reproduce the same (4, 3) simulation as above but replacing the VQD algorithm by the lighter VQE-AC. This is to validate the latter as a viable approach for this molecule with the benefit of not having to set the $\beta$ hyperparameters. For this we have to use the COBYLA optimizer to adjust for the orthogonality constraint, and the UCCSD ansatz as it was the best performing ansatz in this experiment.

    \item \textbf{Experiment 4 -- Further exploring \ce{H2O} deformations for conical intersections, \ce{CH2NH} simulation:} As finding conical intersections in the water molecule is trickier than the methanimine case,  we explore using the classical state-average method a possible minimal configuration for \ce{H2O} that could be used for obtaining conical intersection results using quantum computing methods. We also perform state-average computations for methanimine to compare the results with exact diagonalization results.

\end{itemize}

%%%%%%%%%%%%%%%%%%%%%%%%%%%%%%%%%%%%%%%%%%%%%%%%%%%%%%%%%%%%%%%%%%%%%%%%%%%%%%%%%%%%%%%%%%%%%%%%%%%%%%%%
\section{Results and Discussion}
\label{seq:results}

\textbf{\ce{H2O} -- VQE, VQD, VQE-AC}\\
In this set of simulations, we implemented the 3 variational quantum algorithms for the same \ce{H2O} molecular system: a symmetric geometry where the O-H bond lengths are equal and scaled by a scaling factor d.

\begin{figure}[t]
    \centering
    \includegraphics[width=0.95\linewidth]{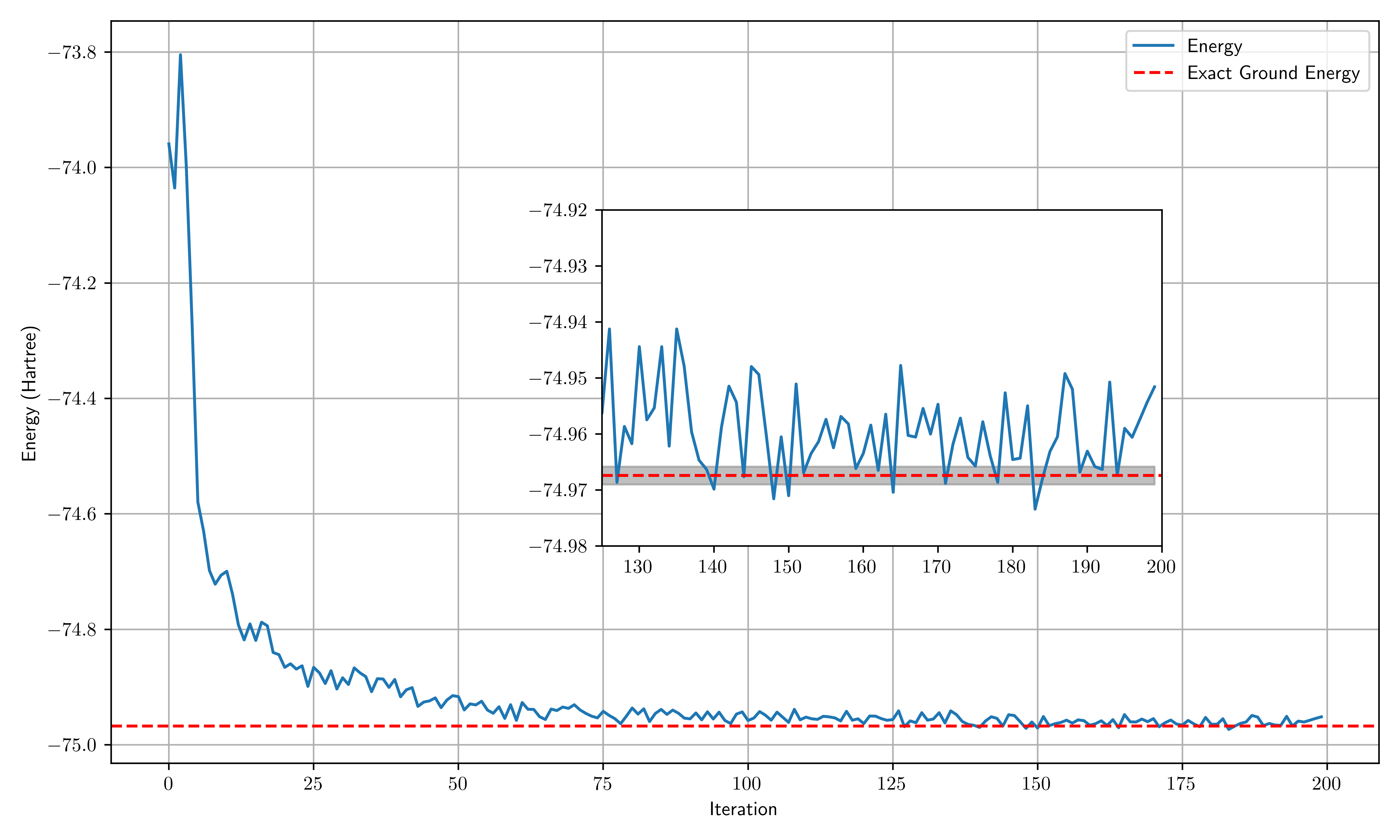}
    \caption{Energy convergence of the VQE algorithm for the ground state energy of the \ce{H2O} molecule as a function of iteration. The red dashed line indicates the exact ground energy in the noiseless case. The inset shows a zoomed-in view of the convergence behavior after 125 iterations.}
    \label{fig:vqe-h2o}
\end{figure}

Fig. \ref{fig:vqe-h2o} shows the convergence of the noiseless VQE algorithm for estimating the ground-state energy of the  molecule (under the considered active space approximations). We observe larger fluctuations in the energy convergence during the initial iterations, particularly between 0 and 10. This behavior is characteristic of the optimizer's initial exploration of the parameter landscape associated with the quantum ansatz. As the optimization process progresses with increasing iterations, the energy generally exhibits a decreasing trend, and the magnitude of these fluctuations diminishes as the SPSA learning rate gets smaller. The VQE ultimately gets closer and closer towards the target value of -74.96742 Hartree. \\

\begin{figure}[t]
    \centering
    \includegraphics[width=0.95\linewidth]{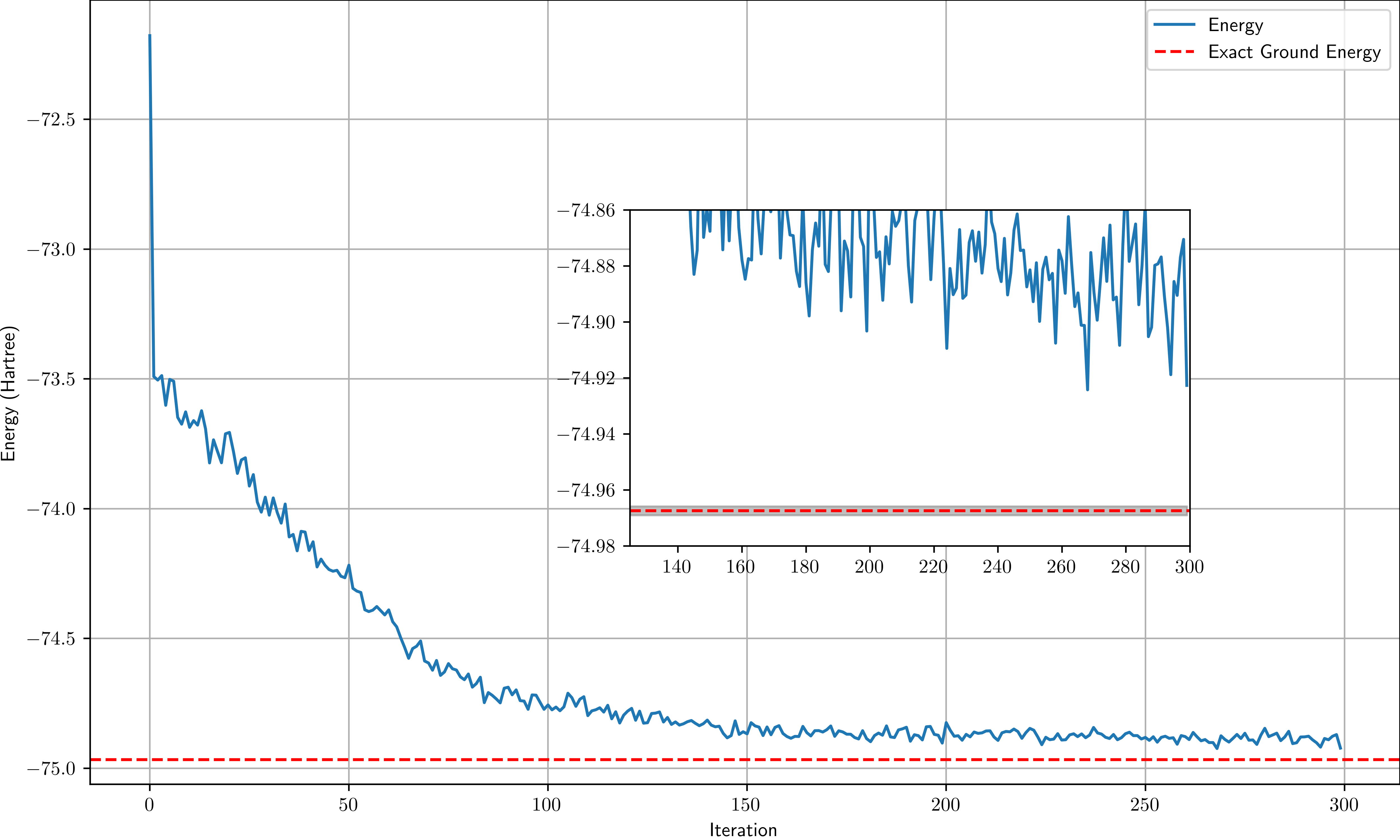}
    \caption{Convergence graph of the VQE algorithm for the ground state energy of the \ce{H2O} molecule as a function of iteration. The red dashed line indicates the exact ground-state energy.}
    \label{fig:vqe-h2o-noisy}
\end{figure}
 
In the noisy case, shown in Fig. \ref{fig:vqe-h2o-noisy}, a similar convergence trend is observed, except that the VQE now converges above the target value. This is a direct result of simulated quantum noise as it prevents an accurate estimation of energies via quantum circuits. Furthermore, the estimated ground energy is now clearly outside the bounds for chemical accuracy.

\begin{figure}[t]
    \centering
    \includegraphics[width=0.95\linewidth]{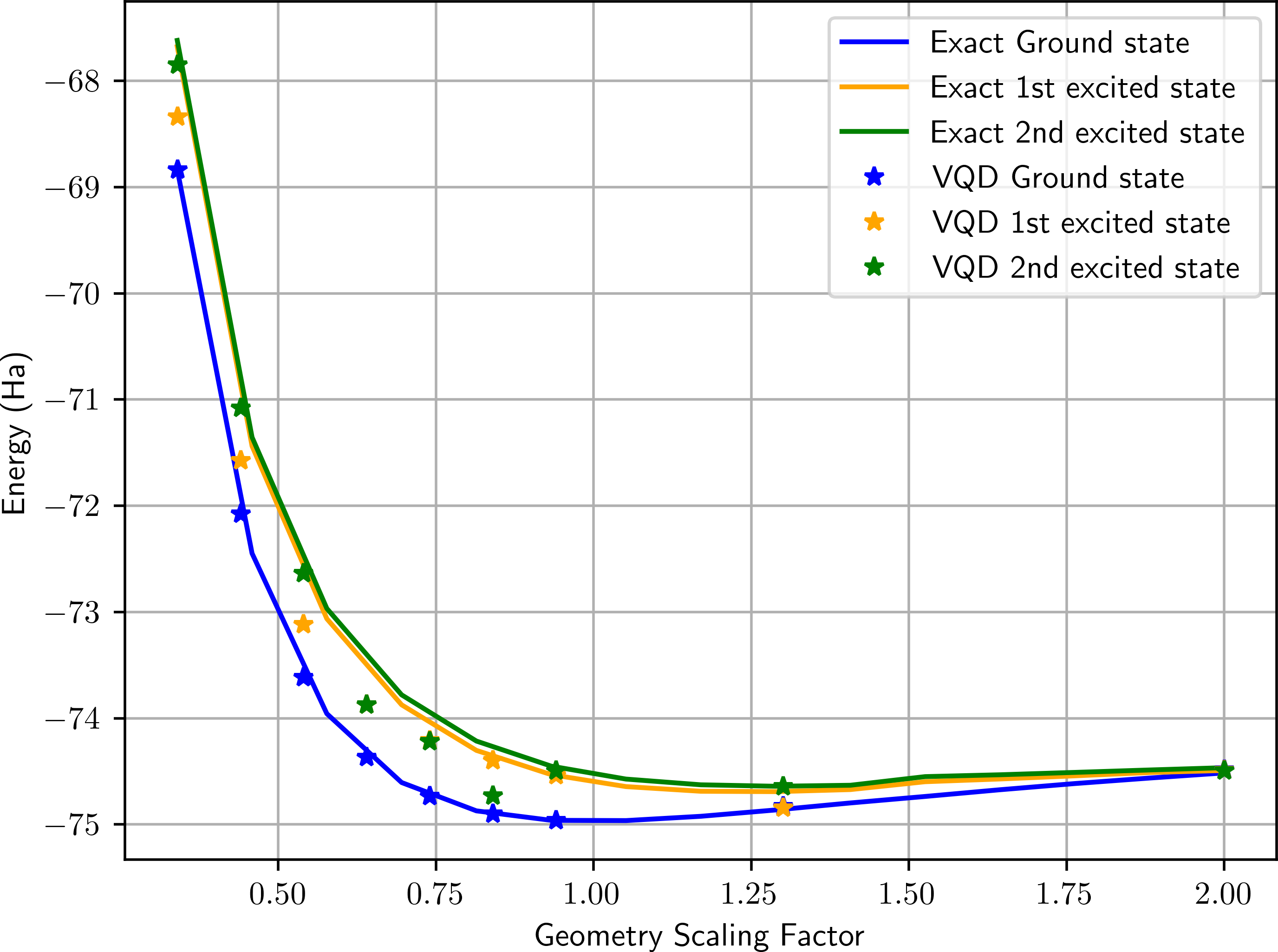}
    \caption{Exact diagonalization versus VQD results for the first 3 energy levels of the \ce{H2O} molecule at scaling factors between 0.34 to 2.}
    \label{fig:vqd-h2o}
\end{figure}

Fig. \ref{fig:vqd-h2o} presents the potential energy curves for the ground and first two excited states of the \ce{H2O} molecule computed using the Variational Quantum Deflation (VQD) method and compared against exact diagonalization results. The VQD energies for the ground state show excellent agreement with exact diagonalization across all geometry scaling factors, confirming the robustness of the method in accurately capturing the lowest energy surface. For the first and second excited states, the VQD results follow the expected qualitative trends but exhibit slight quantitative deviations from the exact values. These deviations are attributed to the limitations of the chosen ansatz and optimization depth, as well as the increasing complexity of excited-state wave functions. These factors chiefly affect the satisfaction of the orthogonality constraint in the VQD, and thus lead to lower estimated energies for the excited states. Satisfying this constraint is further complicated as it enters as a penalty term in the VQD’s cost function, requiring a fine tuning of the $\beta$ hyperparameter that directly affects how the constraint is respected.

\begin{figure}[t]
    \centering
    \includegraphics[width=0.95\linewidth]{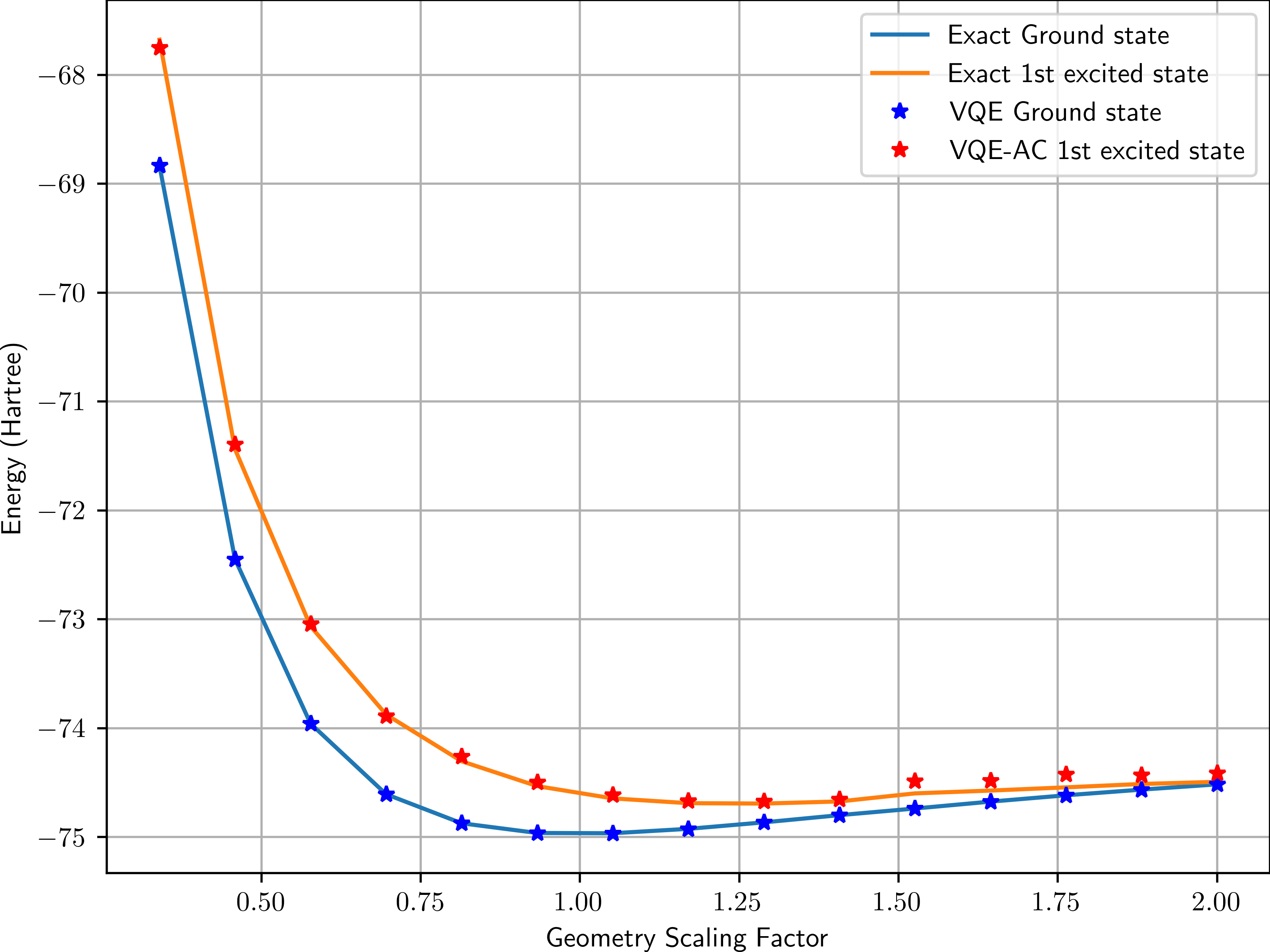}
    \caption{VQE-AC for the \ce{H2O} molecule, solving for the ground- and first-excited state energies. The scaling factor squeezes or stretches the O-H bond distances.}
    \label{fig:vqeac-h2o}
\end{figure}

When switching from the VQD to the VQE-AC algorithm for the same \ce{H2O} problem configuration, we obtain significantly better results that are closer to the exact solution than the VQD results, as is presented in Fig. \ref{fig:vqeac-h2o}. This is partly due to the fact that we did not have to tune an extra hyper parameter, $\beta$, in the VQE-AC. This clearly highlights the main advantage of the latter algorithm over the VQD, which is to simplify the cost function compared to the VQD case, as described in Eqs. \ref{eq:vqe-ac-cost} and \ref{eq:vqe-ac-constraint}.\\

\textbf{\ce{CH2NH} — AVOIDED CROSSING, VQD}\\

Here we present the VQD simulation results at two different active space configurations for the methanimine molecule: 4 electrons and 3 molecular orbitals (4, 3), and 12 electrons and 9 molecular orbitals (12, 9).

\begin{figure}[t]
    \centering
    \includegraphics[width=0.95\linewidth]{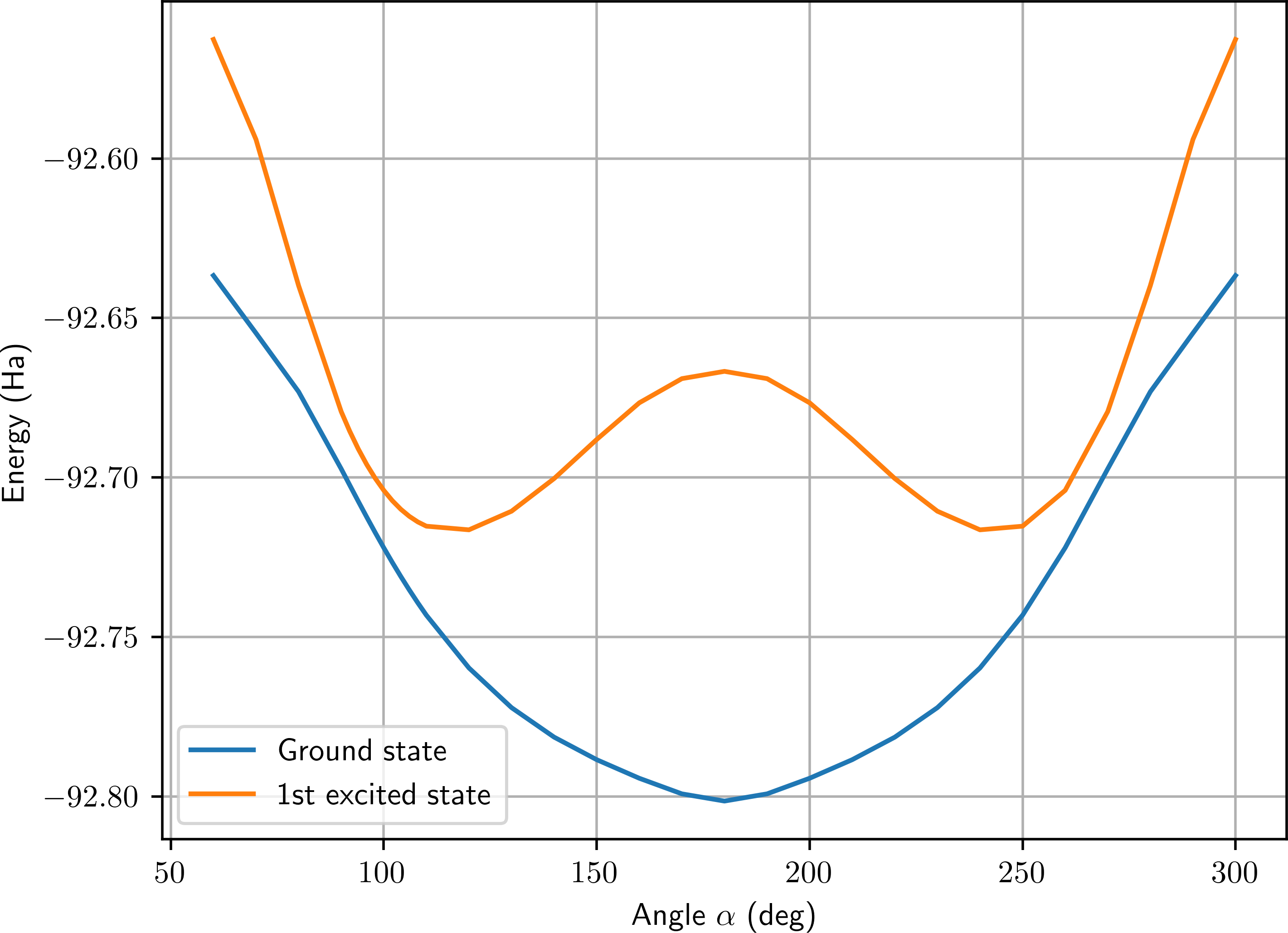}
    \caption{Ground- and first excited-state energies of \ce{CH2NH}, at a (4, 3) active space, and at bending angles $\alpha$ from 60° to 300°. We notice two symmetric avoided crossing points.}
    \label{fig:exact-ch2nh}
\end{figure}

In Fig. \ref{fig:exact-ch2nh} we first show the exact diagonalization curves for the first two energy levels. As a direct consequence from the chosen methanimine geometry, we obtain two symmetric avoided crossing points. It is thus sufficient to study only one of the two regions with the VQD.

\begin{figure}[t]
    \centering
    \includegraphics[width=0.95\linewidth]{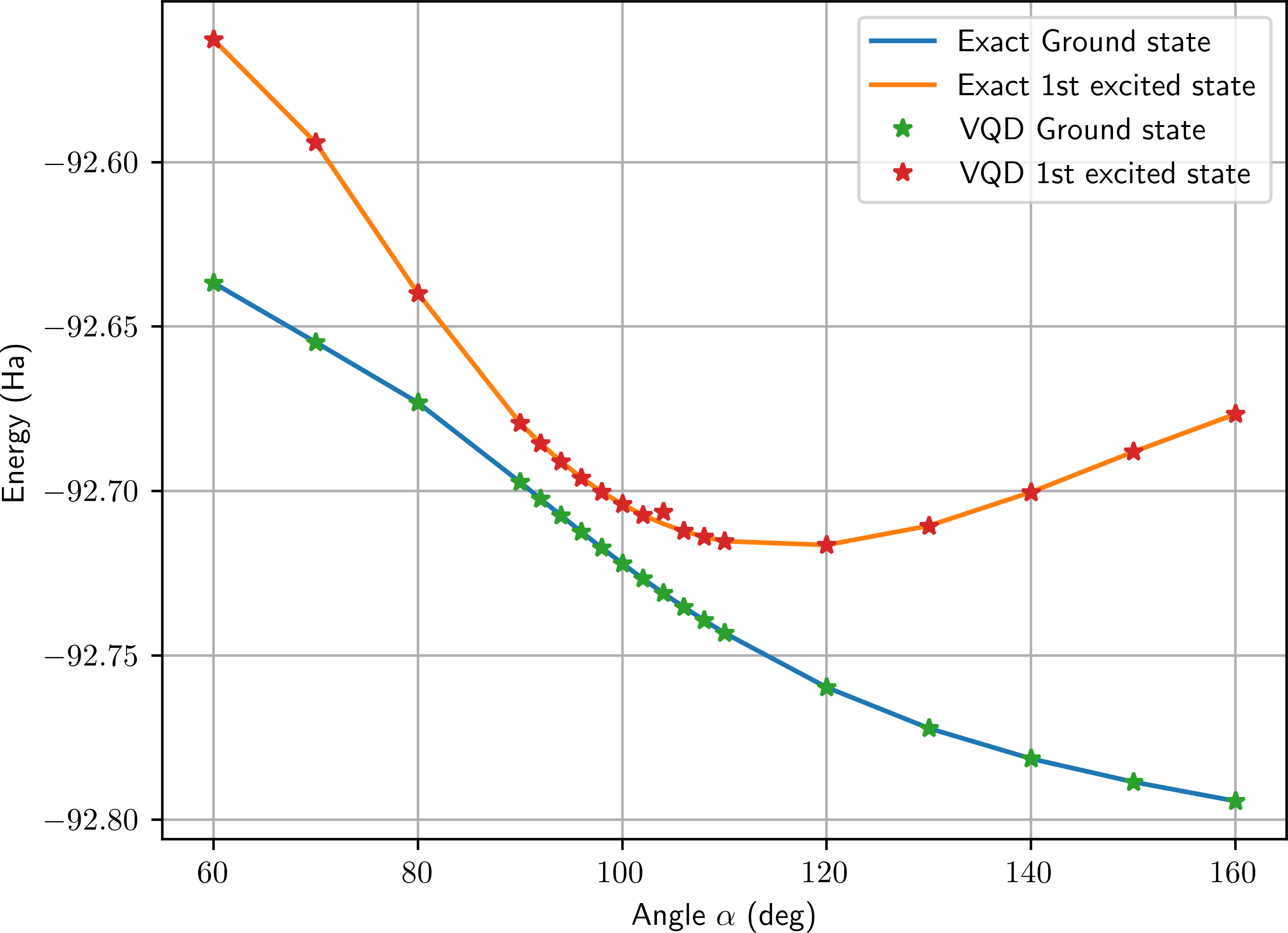}
    \caption{Comparison between the exact diagonalization curves (solid lines) for the ground and the first excited states and the VQD results (stars), with a higher x-resolution around the avoided crossing point.}
    \label{fig:vqd-ch2nh}
\end{figure}

Fig. \ref{fig:vqd-ch2nh} is a comparison between the exact diagonalization curves for both the ground and the first excited states, and the corresponding results obtained from the variational quantum deflation (VQD) algorithm. We chose a higher angle resolution of $2^\circ$ around the avoided conical intersection point, to accurately capture the dynamics and energy crossings between the electronic states. The exact diagonalization solutions serve as a benchmark for the VQD results. This detailed comparison allows for a precise evaluation of the VQD algorithm's accuracy. The VQD results indicate that the avoided crossing point occurs at $\alpha= 94^\circ \pm 2^\circ$, using the (4, 3) active space.
It is also worth noting that the VQD results coincide better with the exact curves for the ground and the first excited states, as opposed to the earlier \ce{H2O} case. We note that the main difference between the two VQDs is that this time we restricted ourselves to only two energy levels, which reduces the errors due to orthogonality constraint.\\

\begin{figure}[t]
    \centering
    \includegraphics[width=0.95\linewidth]{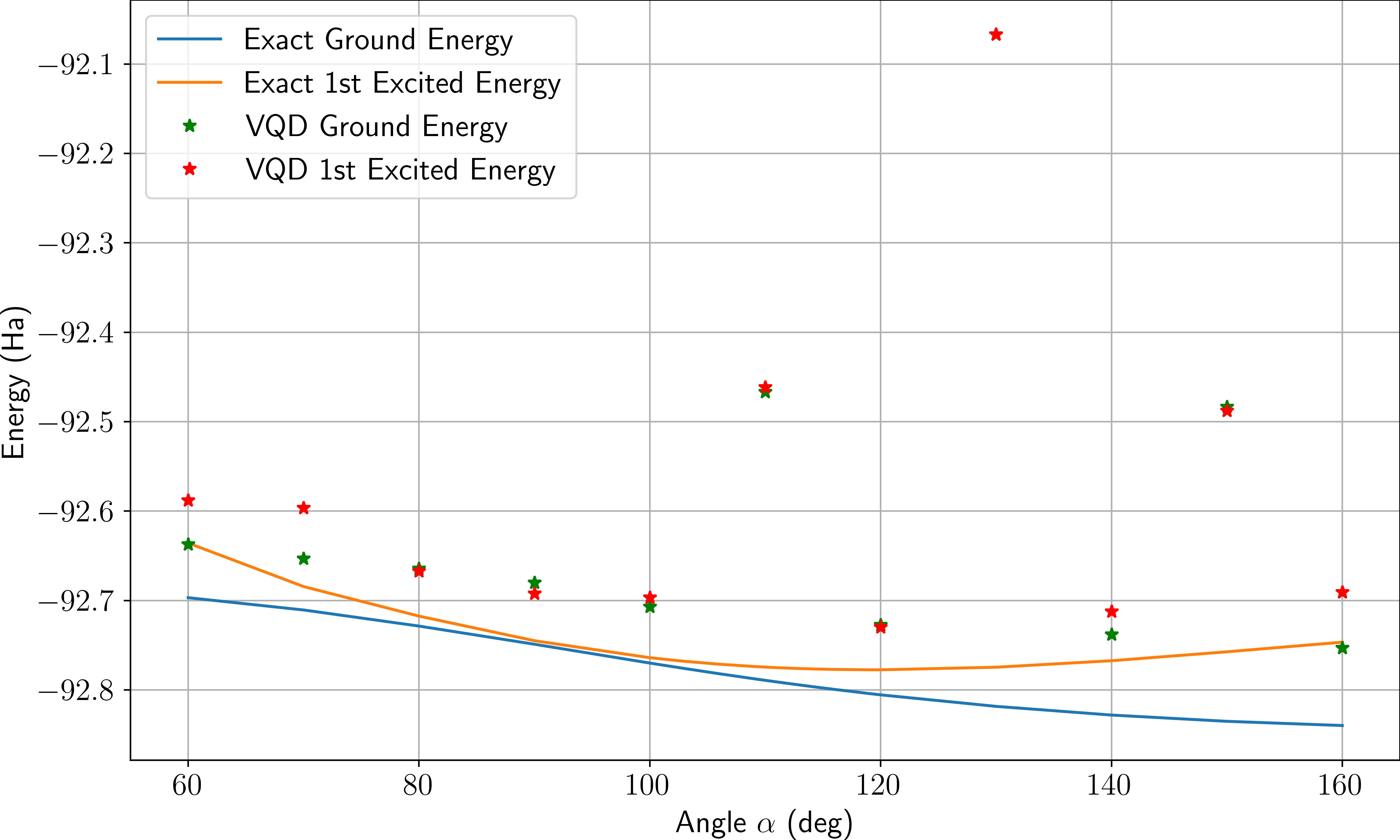}
    \caption{Comparison between the exact diagonalization curves (solid lines) for the ground and the first excited states and the VQD results (stars) for a much larger active space of (12, 9).}
    \label{fig:vqd-ch2nh-12-9}
\end{figure}

In Fig. \ref{fig:vqd-ch2nh-12-9}, we observe that the VQD results for both the ground and first excited states exhibit three distinct local minima. These are artifacts from the optimization procedure where the optimizer is unable to reach the correct ground state that would correspond the the global minimum. 
Importantly, the VQD algorithm, whose results are shown as discrete points, qualitatively replicates the behavior of the exact solutions. It preserves the same overall trend and distinctly illustrates these local minima. However, it is evident that all energy values obtained through VQD are consistently higher when compared to the exact solutions. This upward shift in energy across the entire curve is likely attributable to the limited expressiveness of the chosen ansatz for the (12,9) active space, meaning the quantum circuit might not be sufficiently adequate to fully capture the true ground and excited states' wavefunctions. A significant observation from the comparison is that the gap between the ground state and the first excited state, particularly around the conical intersection point located at around $\alpha =90^\circ \pm 10^\circ$, becomes notably closer as the size of the active space is increased. This behavior underscores the importance of a sufficiently large active space for accurately describing the electronic structure.\\

\textbf{\ce{CH2NH} — AVOIDED CROSSING, VQE-AC}\\

After the VQD simulations, we have carried out a VQE-AC simulation for the methanimine molecule at a (4, 3) active space. We recall here that the ansatz was changed to UCCSD for better results in this case.

\begin{figure}[t]
    \centering
    \includegraphics[width=0.95\linewidth]{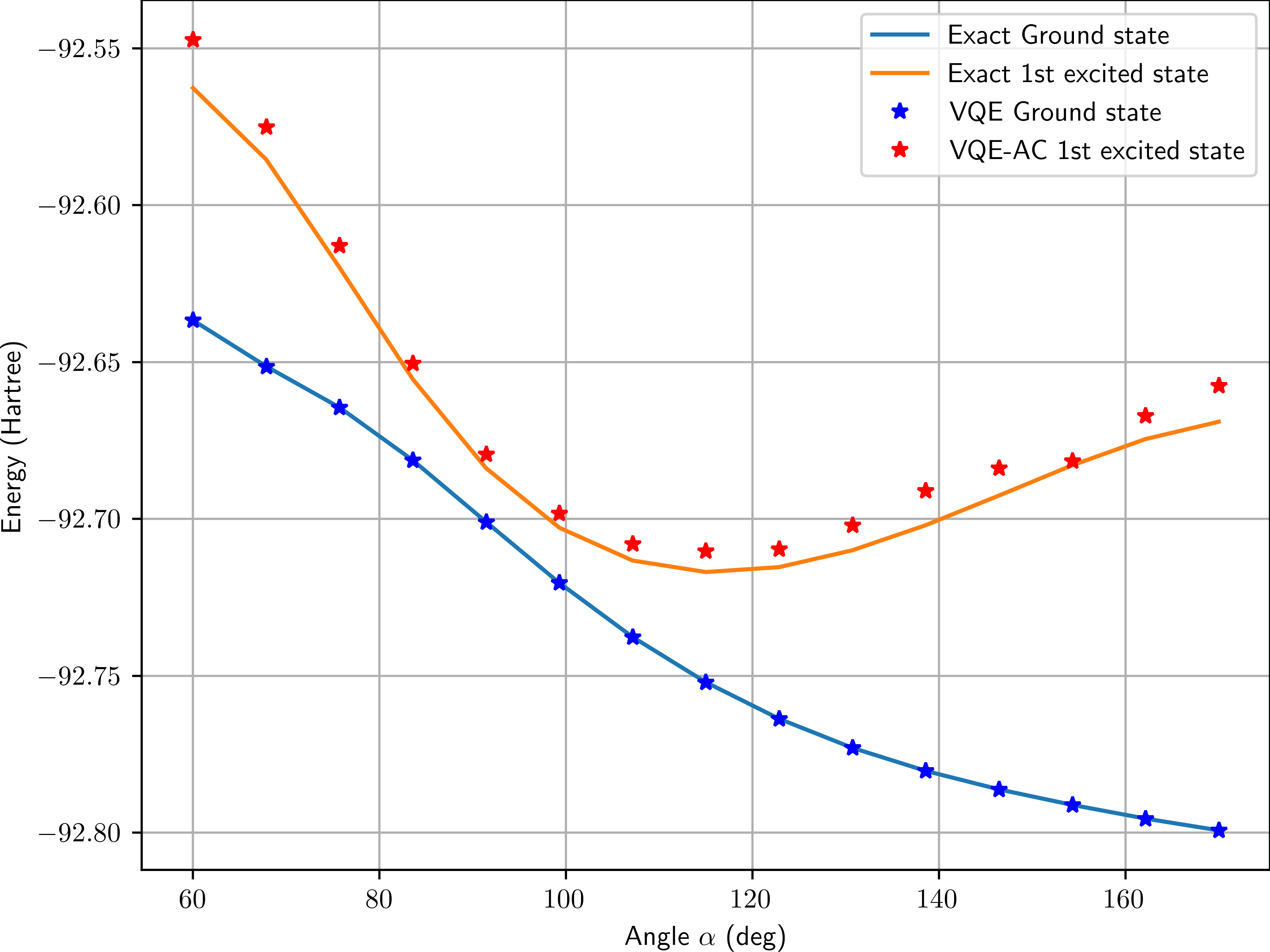}
    \caption{Energy curves for \ce{CH2NH} as a function of the bending angle $\alpha$. The solid blue and orange lines are the exact ground- and first excited-state energies, respectively. The blue and red stars are the VQE-AC ground- (VQE) and first excited-state (VQE-AC) energies.}
    \label{fig:ch2nh-vqe-ac}
\end{figure}

Using the variational quantum eigensolver with automatically-adjusted constraints algorithm to calculate the electronic structure of the \ce{CH2NH} molecule, as evident from the plotted data in Fig. \ref{fig:ch2nh-vqe-ac}, we obtained a close agreement between the VQE-AC results—with its initial VQE and the subsequent VQE-AC components—and their respective exact diagonalization curves, particularly for the ground state. This result confirms the efficacy of this approach for accurately determining molecular energies for the first two energy levels. In addition, this shows that the VQE-AC can be a simpler replacement for the VQD, provided that one makes the necessary adjustments to the ansatz (from Efficient SU2 to UCCSD in this case). Since the molecular system is the same as the earlier one, the avoided crossing point that was obtained via the VQE-AC stayed located around the same $\alpha=94^\circ$ region.\\

\textbf{SA — Exploring further deformations for \ce{H2O}, application on \ce{CH2NH}}\\

We now use the state-average CASSCF method for further explorations related to both molecules. Indeed for the water molecule, these results serve as a guide toward choosing configurations that allows for the finding of conical intersections.

\begin{figure}[t]
    \centering
    \includegraphics[width=0.95\linewidth]{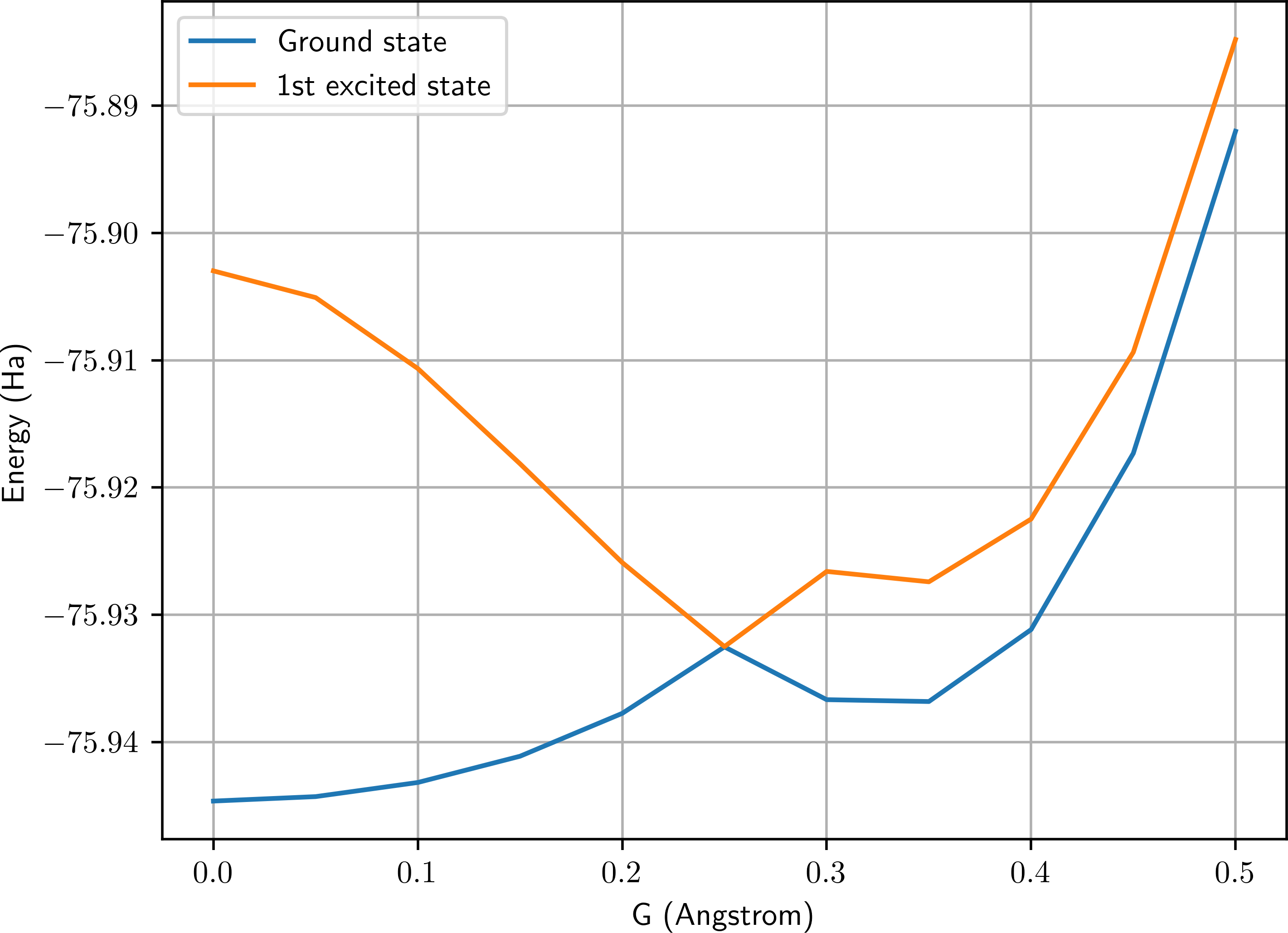}
    \caption{State-average simulation for a (10, 12) active space and 6-31G basis set \ce{H2O} system. Results were obtained using PySCF \cite{Sun2020}.}
    \label{fig:state-h2o}
\end{figure}

Here the energy curves for the ground- and first excited-states of the water molecule were calculated for the asymmetric deformation described in subsection \ref{subseq:exp-setup}. As shown in Figure \ref{fig:state-h2o}, the potential energy surfaces of the two states intersect each other around $G=0.25~\AA$, (we recall that $\gamma$ is fixed to $0.00021^\circ$) indicating the presence of a conical intersection (CI) in that region. At this geometry, the energy difference between the two states vanishes. Furthermore, the PES in this region shows a high degree of sensitivity to small changes to the system’s Hamiltonian. Changes to the basis set or the active space could make the CI disappear in the SA results. This highlights the complex topology of the electronic structure landscape near the CI. In contrast, when considering the O-H symmetry as was the case in earlier simulations, no intersection could be found. These symmetric geometries, even in the case of large active spaces, are insufficient for describing true crossings or conical intersections for the water molecule. The requirement of the (10, 12) active space and 6-31G basis set significantly increase the quantum computational cost for uncovering CIs in this case. Simulations of quantum algorithms applied on these larger systems, which would require significant classical computational resources, could be addressed in future works. However, implementations on real quantum computers for the simulation of conical intersections in the water molecule should be within the current capabilities of NISQ era devices, as long as appropriate optimizations and error mitigation schemes are utilized to counteract the effects of noise.

\begin{figure}[t]
    \centering
    \includegraphics[width=0.95\linewidth]{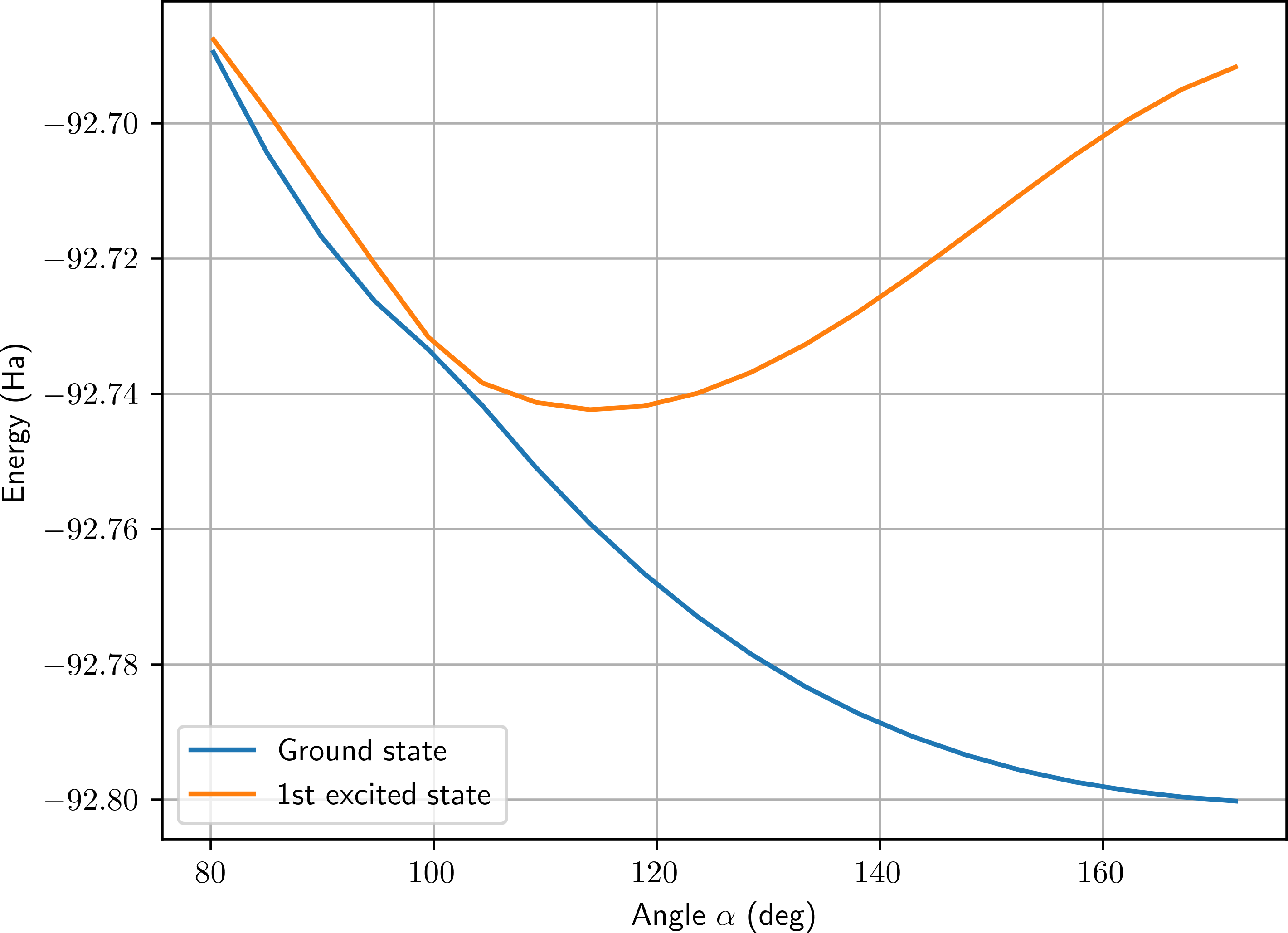}
    \caption{State-average results for \ce{CH2NH} with a (4, 3) active space.}
    \label{fig:state-ch2nh}
\end{figure}

The potential energy curves of the ground and first excited states of the methanimine molecule were re-computed using state-average CASSCF method with a (4, 3) active space. Fig. \ref{fig:state-ch2nh} qualitatively reproduces the earlier exact diagonalization curves presented above. This time the avoided crossing point is situated around $\alpha = 100^\circ$. The near-degeneracy of the electronic states obtained again here is indicative of a conical intersection. The state-average approach used here is crucial for correctly capturing this intersection at a lower cost than the exact diagonalization method. This means that for benchmarking quantum algorithms in our use cases one might consider this cheaper method instead, especially since we are chiefly interested in qualitative behaviors of the energy curves to serve as indications for where to apply the more expensive quantum methods.

%%%%%%%%%%%%%%%%%%%%%%%%%%%%%%%%%%%%%%%%%%%%%%%%%%%%%%%%%%%%%%%%%%%%%%%%%%%%%%%%%%%%%%%%%%%%%%%%%%%%%%%%
\section{Conclusion}
\label{seq:conclusion}
In this study, we employed variational quantum algorithms (VQAs) to compute the ground and excited state energies of water (\ce{H2O}) and methanimine (\ce{CH2NH}), with the goal of identifying and characterizing conical intersections. Using the Variational Quantum Eigensolver (VQE), along with its excited-state extensions---Variational Quantum Deflation (VQD) and VQE with Automatically-Adjusted Constraints (VQE-AC)---as well as the classical State-Average (SA) method, we evaluated the effectiveness of these techniques under varying molecular geometries, basis sets, active spaces, and ansätze. Our results confirm the reliability of VQE for ground-state estimation, even in the presence of simulated noise. For excited states, VQD and VQE-AC both provided accurate representations of the energy landscape, particularly near regions of strong state coupling. The VQE-AC method proved especially advantageous in sensitive regions such as those containing conical intersections by enforcing orthogonality constraints without requiring manual tuning of penalty parameters. The classical state-average method offered valuable reference data and helped guide the identification of intersection regions, especially when large active spaces are considered. For the \ce{CH2NH} molecule, an avoided crossing point, indicative of a conical intersection, was located in a region of bending angles, $\alpha$, centered around $90^\circ$ to $100^\circ$ within the chosen geometry, and depending on the active space size. While for \ce{H2O}, an intersection was observed at the Jacobi coordinate of $G=0.25 \AA$, also within its own deformed geometry. Notably, this feature in water could not be resolved when using more symmetric geometries, emphasizing the importance of asymmetric deformations in capturing such effects, and using internal coordinate systems for easier modelization. Furthermore, we found that larger active spaces significantly improved the precision and expressiveness of quantum simulations in the vicinity of conical intersections. This however comes at an additional computational cost, and with challenges related to choosing appropriate trial states. This work demonstrates the viability of near-term quantum algorithms in modeling complex electronic structures and excited-state phenomena. Our findings underline the importance of computational choices (geometry, ansatz, optimizer, and active space) in obtaining accurate quantum chemical predictions, and suggest promising future directions for quantum simulations in strongly correlated molecular systems.

%%%%%%%%%%%%%%%%%%%%%%%%%%%%%%%%%%%%%%%%%%%%%%%%%%%%%%%%%%%%%%%%%%%%%%%%%%%%%%%%%%%%%%%%%%%%%%%%%%%%%%%%
\section*{Acknowledgments}
This document has been produced with the financial assistance of the European Union (Grant no. DCI-PANAF/2020/420-028), through the African Research Initiative for Scientific Excellence (ARISE), pilot programme. ARISE is implemented by the African Academy of Sciences with support from the European Commission and the African Union Commission. The contents of this document are the sole responsibility of the author(s) and can under no circumstances be regarded as reflecting the position of the European Union, the African Academy of Sciences, and the African Union Commission. We are grateful to Dr. Mohamed Taha Rouabah for acquiring funding through the ARISE pilot programme grant, problem conceptualization, his management and administrative work that made this work possible, and for his validation of the manuscript. We express our deepest gratitude to the Algerian Ministry of Higher Education and Scientific Research (MESRS) and the General Directorate for Scientific Research and Technological Development (DGRSDT) for their financial support. We are also indebted to Professor Nadir Belaloui for his insightful discussions throughout this work.

\bibliography{references}

\begin{thebibliography}{62}
\providecommand{\natexlab}[1]{#1}
\providecommand{\url}[1]{\texttt{#1}}
\expandafter\ifx\csname urlstyle\endcsname\relax
  \providecommand{\doi}[1]{doi: #1}\else
  \providecommand{\doi}{doi: \begingroup \urlstyle{rm}\Url}\fi

\bibitem[Nielsen and Chuang(2010)]{Nielsen2010}
Michael~A. Nielsen and Isaac~L. Chuang.
\newblock \emph{Quantum Computation and Quantum Information}.
\newblock Cambridge University Press, Cambridge, 10th anniversary edition edition, 2010.

\bibitem[Feynman(1982)]{Feynman1982}
Richard~P. Feynman.
\newblock Simulating physics with computers.
\newblock \emph{International Journal of Theoretical Physics}, 21\penalty0 (6-7):\penalty0 467--488, 1982.
\newblock \doi{10.1007/BF02650179}.

\bibitem[Benioff(1980)]{Benioff1980}
Paul Benioff.
\newblock The computer as a physical system: A microscopic quantum mechanical hamiltonian model of computers as represented by turing machines.
\newblock \emph{Journal of Statistical Physics}, 22\penalty0 (5):\penalty0 563--591, 1980.
\newblock \doi{10.1007/BF01011339}.

\bibitem[Manin(1980)]{Manin1980}
Yuri~I. Manin.
\newblock \emph{Computable and Uncomputable}.
\newblock Sovetskoye Radio, Moscow, 1980.
\newblock in Russian.

\bibitem[Kassal et~al.(2011)Kassal, Whitfield, Perdomo-Ortiz, Yung, and Aspuru-Guzik]{Kassal2011}
Ivan Kassal, James~D. Whitfield, Alejandro Perdomo-Ortiz, Man-Hong Yung, and Al{\'a}n Aspuru-Guzik.
\newblock Simulating chemistry using quantum computers.
\newblock \emph{Annual Review of Physical Chemistry}, 62:\penalty0 185--207, 2011.
\newblock \doi{10.1146/annurev-physchem-032210-103512}.

\bibitem[Cai et~al.(2020)Cai, Fang, Fan, and Li]{Cai2020}
Xiaoxia Cai, Wei-Hai Fang, Heng Fan, and Zhendong Li.
\newblock Quantum computation of molecular response properties.
\newblock \emph{Phys. Rev. Res.}, 2:\penalty0 033324, Aug 2020.
\newblock \doi{10.1103/PhysRevResearch.2.033324}.
\newblock URL \url{https://link.aps.org/doi/10.1103/PhysRevResearch.2.033324}.

\bibitem[Hartree(1928)]{Hartree1928}
D.~R. Hartree.
\newblock The wave mechanics of an atom with a non-coulomb central field. part i. theory and methods.
\newblock \emph{Mathematical Proceedings of the Cambridge Philosophical Society}, 24\penalty0 (1):\penalty0 89--110, 1928.
\newblock \doi{10.1017/S0305004100011919}.

\bibitem[Fock(1930)]{Fock1930}
V.~A. Fock.
\newblock N{\"a}herungsmethode zur {L}{\"o}sung des quantenmechanischen mehrk{\"o}rperproblems.
\newblock \emph{Zeitschrift f{\"u}r Physik}, 61:\penalty0 126--148, 1930.
\newblock \doi{10.1007/BF01340294}.

\bibitem[Kohn(1999)]{Kohn1999}
Walter Kohn.
\newblock Nobel lecture: Electronic structure of matter---wave functions and density functionals.
\newblock \emph{Reviews of Modern Physics}, 71:\penalty0 1253--1266, 1999.
\newblock \doi{10.1103/RevModPhys.71.1253}.

\bibitem[White(1992)]{White1992}
Steven~R. White.
\newblock Density matrix formulation for quantum renormalization groups.
\newblock \emph{Physical Review Letters}, 69\penalty0 (19):\penalty0 2863--2866, 1992.
\newblock \doi{10.1103/PhysRevLett.69.2863}.

\bibitem[White(1993)]{White1993}
Steven~R. White.
\newblock Density-matrix algorithms for quantum renormalization groups.
\newblock \emph{Physical Review B}, 48\penalty0 (14):\penalty0 10345--10356, 1993.
\newblock \doi{10.1103/PhysRevB.48.10345}.

\bibitem[Shavitt(1981)]{Shavitt1981}
Isaiah Shavitt.
\newblock The method of configuration interaction.
\newblock In David~R. Yarkony, editor, \emph{Methods of Electronic Structure Theory, Vol. 2}, pages 189--275. Academic Press, New York, 1981.

\bibitem[McArdle et~al.(2020)McArdle, Endo, Aspuru-Guzik, Benjamin, and Yuan]{McArdle2020}
Sam McArdle, Suguru Endo, Al{\'a}n Aspuru-Guzik, Simon~C. Benjamin, and Xiao Yuan.
\newblock Quantum computational chemistry.
\newblock \emph{Reviews of Modern Physics}, 92\penalty0 (1):\penalty0 015003, 2020.
\newblock \doi{10.1103/RevModPhys.92.015003}.

\bibitem[Neese(2017)]{neese2017}
Frank Neese.
\newblock High-level spectroscopy, quantum chemistry, and catalysis: Not just a passing fad.
\newblock \emph{Angewandte Chemie International Edition}, 56\penalty0 (37):\penalty0 11003--11010, 2017.

\bibitem[Shokrian~Zini et~al.(2023)Shokrian~Zini, Delgado, dos Reis, Moreno~Casares, Mueller, Voigt, and Arrazola]{Zini2023}
Modjtaba Shokrian~Zini, Alain Delgado, Roberto dos Reis, Pablo~Antonio Moreno~Casares, Jonathan~E. Mueller, Arne-Christian Voigt, and Juan~Miguel Arrazola.
\newblock Quantum simulation of battery materials using ionic pseudopotentials.
\newblock \emph{Quantum}, 7:\penalty0 1049, 2023.
\newblock \doi{10.22331/q-2023-07-10-1049}.

\bibitem[Gross and Marques(2012)]{Gross2012}
Eberhard K.~U. Gross and Miguel A.~L. Marques, editors.
\newblock \emph{Fundamentals of Time-Dependent Density Functional Theory}, volume 837 of \emph{Lecture Notes in Physics}.
\newblock Springer, Berlin, Heidelberg, 2012.

\bibitem[Peruzzo et~al.(2014)Peruzzo, McClean, Shadbolt, Yung, Zhou, Love, Aspuru-Guzik, and O'Brien]{Peruzzo2014}
Alberto Peruzzo, Jarrod McClean, Peter Shadbolt, Man-Hong Yung, Xiao-Qi Zhou, Peter~J. Love, Al{\'a}n Aspuru-Guzik, and Jeremy~L. O'Brien.
\newblock A variational eigenvalue solver on a photonic quantum processor.
\newblock \emph{Nature Communications}, 5:\penalty0 4213, 2014.
\newblock \doi{10.1038/ncomms5213}.

\bibitem[Tilly et~al.(2022)Tilly, Chen, Cao, Picozzi, Setia, Li, Grant, Wossnig, Rungger, Booth, and Tennyson]{Tilly2022}
Jules Tilly, Hongxiang Chen, Shuxiang Cao, Dario Picozzi, Kanav Setia, Ying Li, Edward Grant, Leonard Wossnig, Ivan Rungger, George~H. Booth, and Jonathan Tennyson.
\newblock The variational quantum eigensolver: a review of methods and best practices.
\newblock \emph{Physics Reports}, 986:\penalty0 1--128, 2022.
\newblock \doi{10.1016/j.physrep.2022.08.003}.

\bibitem[Cohen-Tannoudji et~al.(2017)Cohen-Tannoudji, Diu, and Lalo{\"e}]{Cohen-Tannoudji2017}
Claude Cohen-Tannoudji, Bernard Diu, and Franck Lalo{\"e}.
\newblock \emph{M{\'e}canique quantique - Tome III}.
\newblock EDP Sciences, 2017.

\bibitem[Kitaev(1995)]{Kitaev1995}
A.~Y. Kitaev.
\newblock Quantum measurements and the abelian stabilizer problem.
\newblock \emph{arXiv preprint arXiv:quant-ph/9511026}, 1995.

\bibitem[Motta et~al.(2020)Motta, Sun, Tan, O'Rourke, Ye, Minnich, Brand{\~a}o, and Chan]{Motta2020}
Mario Motta, Chong Sun, Adrian T.~K. Tan, Matthew~J. O'Rourke, Erika Ye, Austin~J. Minnich, Fernando G. S.~L. Brand{\~a}o, and Garnet K.-L. Chan.
\newblock Quantum imaginary time evolution, quantum lanczos, and variational quantum algorithms.
\newblock \emph{Quantum}, 4:\penalty0 314, 2020.
\newblock \doi{10.22331/q-2020-05-25-314}.

\bibitem[Farhi et~al.(2000)Farhi, Goldstone, Gutmann, and Sipser]{Farhi2000}
Edward Farhi, Jeffrey Goldstone, Sam Gutmann, and Michael Sipser.
\newblock Quantum computation by adiabatic evolution.
\newblock \emph{arXiv preprint arXiv:quant-ph/0001106}, 2000.

\bibitem[Roothaan(1951)]{Roothaan1951}
C.~C.~J. Roothaan.
\newblock New developments in molecular orbital theory.
\newblock \emph{Reviews of Modern Physics}, 23\penalty0 (2):\penalty0 69--89, 1951.
\newblock \doi{10.1103/RevModPhys.23.69}.

\bibitem[Stanton and Bartlett(1993)]{Stanton1993}
John~F. Stanton and Rodney~J. Bartlett.
\newblock The equation of motion coupled-cluster method: A systematic biorthogonal approach to molecular excitation energies, transition probabilities, and excited state properties.
\newblock \emph{The Journal of Chemical Physics}, 98\penalty0 (9):\penalty0 7029--7039, 1993.
\newblock \doi{10.1063/1.464746}.

\bibitem[Runge and Gross(1984)]{Runge1984}
Erich Runge and E.~K.~U. Gross.
\newblock Density-functional theory for time-dependent systems.
\newblock \emph{Physical Review Letters}, 52\penalty0 (12):\penalty0 997--1000, 1984.
\newblock \doi{10.1103/PhysRevLett.52.997}.

\bibitem[Roos et~al.(1980)Roos, Taylor, and Siegbahn]{Roos1980}
B.~O. Roos, P.~R. Taylor, and P.~E.~M. Siegbahn.
\newblock A complete active space scf method (casscf) using a density matrix formulated super-ci approach.
\newblock \emph{Chemical Physics}, 48\penalty0 (2):\penalty0 157--173, 1980.
\newblock \doi{10.1016/0301-0104(80)80045-0}.

\bibitem[Werner and Knowles(1988)]{Werner1988}
Hans-Joachim Werner and Peter~J. Knowles.
\newblock An efficient internally contracted multiconfiguration–reference configuration interaction method.
\newblock \emph{The Journal of Chemical Physics}, 89\penalty0 (9):\penalty0 5803--5814, 1988.
\newblock \doi{10.1063/1.455556}.

\bibitem[Grimsley et~al.(2019)Grimsley, Economou, Barnes, and Mayhall]{Grimsley2019}
H.~R. Grimsley, Sophia~E. Economou, Edwin Barnes, and Nicholas~J. Mayhall.
\newblock An adaptive variational algorithm for exact molecular simulations on a quantum computer.
\newblock \emph{Nature Communications}, 10:\penalty0 3007, 2019.
\newblock \doi{10.1038/s41467-019-10988-2}.

\bibitem[McClean et~al.(2017)McClean, Kimchi-Schwartz, Carter, and de~Jong]{McClean2017}
Jarrod~R. McClean, Mollie~E. Kimchi-Schwartz, Jonathan Carter, and Wibe~A. de~Jong.
\newblock Hybrid quantum-classical hierarchy for mitigation of decoherence and determination of excited states.
\newblock \emph{Physical Review A}, 95:\penalty0 042308, 2017.
\newblock \doi{10.1103/PhysRevA.95.042308}.

\bibitem[Yalouz et~al.(2021)Yalouz, Senjean, G{\"u}nther, Buda, O'Brien, and Visscher]{Yalouz2021}
Saad Yalouz, Bruno Senjean, Jakob G{\"u}nther, Francesco Buda, Thomas~E. O'Brien, and Lucas Visscher.
\newblock A state-averaged orbital-optimized hybrid quantum--classical algorithm for a democratic description of ground and excited states.
\newblock \emph{Quantum Science and Technology}, 6\penalty0 (2):\penalty0 1--20, 2021.
\newblock \doi{10.1088/2058-9565/abd334}.

\bibitem[Ollitrault et~al.(2020)Ollitrault, Kandala, Chen, Barkoutsos, Mezzacapo, Pistoia, Sheldon, Woerner, Gambetta, and Tavernelli]{Ollitrault2020}
Pauline~J. Ollitrault, Abhinav Kandala, Chun-Fu Chen, Panagiotis~Kl. Barkoutsos, Antonio Mezzacapo, Marco Pistoia, Sarah Sheldon, Stefan Woerner, Jay Gambetta, and Ivano Tavernelli.
\newblock Quantum equation of motion for computing molecular excitation energies on a noisy quantum processor.
\newblock \emph{Physical Review Research}, 2\penalty0 (4):\penalty0 043140, 2020.
\newblock \doi{10.1103/PhysRevResearch.2.043140}.

\bibitem[Cadi~Tazi and Thom(2024)]{cadi2024}
Lila Cadi~Tazi and Alex~JW Thom.
\newblock Folded spectrum vqe: A quantum computing method for the calculation of molecular excited states.
\newblock \emph{Journal of Chemical Theory and Computation}, 20\penalty0 (6):\penalty0 2491--2504, 2024.

\bibitem[Gocho et~al.(2023)Gocho, Nakamura, Kanno, Gao, Kobayashi, Inagaki, Hatanaka, Yagai, Tsukamoto, Harabuchi, Yamato, Meguro, Tanaka, and Watanabe]{Gocho2023}
Sumio Gocho, Hideaki Nakamura, Sho Kanno, Qi~Gao, Toru Kobayashi, Taketo Inagaki, Mai Hatanaka, Shoko Yagai, Hiroyuki Tsukamoto, Yuta Harabuchi, Saori Yamato, Yuka Meguro, Kouyama Tanaka, and Hiroya Watanabe.
\newblock Excited state calculations using variational quantum eigensolver with spin-restricted ans{\"a}tze and automatically-adjusted constraints.
\newblock \emph{npj Computational Materials}, 9:\penalty0 13, 2023.
\newblock \doi{10.1038/s41524-023-00965-1}.

\bibitem[Gonon(2017)]{Gonon2017}
Benjamin Gonon.
\newblock {Simulations quantiques non-adiabatiques d'un photo-interrupteur mol{\'e}culaire vers un dialogue exp{\'e}rience-th{\'e}orie}.
\newblock Ph.D. thesis, Universit{\'e} Montpellier, November 2017.

\bibitem[Yarkony(1996)]{Yarkony1996}
David~R. Yarkony.
\newblock Diabolical conical intersections.
\newblock \emph{Reviews of Modern Physics}, 68:\penalty0 985--1013, 1996.
\newblock \doi{10.1103/RevModPhys.68.985}.

\bibitem[Levine and Martinez(2007)]{Levine2007}
Benjamin~G. Levine and Todd~J. Martinez.
\newblock Isomerization through conical intersections.
\newblock \emph{Annual Review of Physical Chemistry}, 58:\penalty0 613--634, 2007.
\newblock \doi{10.1146/annurev.physchem.57.032905.104612}.

\bibitem[Rivera et~al.(2021)Rivera, Stojanovic, and Crespo-Otero]{Rivera2021}
Miguel Rivera, Ljiljana Stojanovic, and Rachel Crespo-Otero.
\newblock Role of conical intersections on the efficiency of fluorescent organic molecular crystals.
\newblock \emph{The Journal of Physical Chemistry A}, 125\penalty0 (4):\penalty0 1012--1024, 2021.
\newblock \doi{10.1021/acs.jpca.0c11072}.

\bibitem[Zhou et~al.(2023)Zhou, Shi, Long, Yao, Ma, Chen, Du, Sun, Fan, Liu, Wang, Chen, Sui, Yuan, and Peng]{Zhou2023}
Xiao Zhou, Chao Shi, Saran Long, Qichao Yao, He~Ma, Kele Chen, Jianjun Du, Wen Sun, Jiangli Fan, Bin Liu, Lei Wang, Xiaoqiang Chen, Laizhi Sui, Kaijun Yuan, and Xiaojun Peng.
\newblock Highly efficient photosensitizers with molecular vibrational torsion for cancer photodynamic therapy.
\newblock \emph{ACS Central Science}, 9\penalty0 (8):\penalty0 1679--1691, 2023.
\newblock \doi{10.1021/acscentsci.3c00611}.

\bibitem[Szabo and Ostlund(1996)]{Szabo1996}
Attila Szabo and Neil~S. Ostlund.
\newblock \emph{Modern Quantum Chemistry: Introduction to Advanced Electronic Structure Theory}.
\newblock Dover Publications, Mineola, NY, 1996.

\bibitem[Levine(1999)]{Levine1999}
Ira~N Levine.
\newblock \emph{Quantum Chemistry}.
\newblock Pearson, Upper Saddle River, NJ, 5 edition, July 1999.

\bibitem[Javadi-Abhari et~al.(2024)Javadi-Abhari, Treinish, Krsulich, Wood, Lishman, Gacon, Martiel, Nation, Bishop, Cross, Johnson, and Gambetta]{qiskit}
Ali Javadi-Abhari, Matthew Treinish, Kevin Krsulich, Christopher~J. Wood, Jake Lishman, Julien Gacon, Simon Martiel, Paul~D. Nation, Lev~S. Bishop, Andrew~W. Cross, Blake~R. Johnson, and Jay~M. Gambetta.
\newblock Quantum computing with {Q}iskit.
\newblock \emph{arXiv}, page arXiv:2405.08810, May 2024.
\newblock \doi{10.48550/arXiv.2405.08810}.
\newblock URL \url{https://doi.org/10.48550/arXiv.2405.08810}.

\bibitem[Sun et~al.(2020)Sun, Zhang, Banerjee, Bao, Barbry, Bl{\v{a}}{\.z}ek, and Chan]{Sun2020}
Qiming Sun, Xizhou Zhang, Sandeep Banerjee, Pengbo Bao, Murat Barbry, Jakub Bl{\v{a}}{\.z}ek, and Garnet K.-L. Chan.
\newblock Recent developments in the pyscf program package.
\newblock \emph{The Journal of Chemical Physics}, 153\penalty0 (2):\penalty0 024109, 2020.
\newblock \doi{10.1063/5.0006074}.

\bibitem[Worth and Cederbaum(2004)]{Worth2004}
Graham~A. Worth and Lorenz~S. Cederbaum.
\newblock Beyond born-oppenheimer: Molecular dynamics through a conical intersection.
\newblock \emph{Annual Review of Physical Chemistry}, 55:\penalty0 127--158, 2004.
\newblock \doi{10.1146/annurev.physchem.55.091602.094335}.

\bibitem[Born and Oppenheimer(1927)]{BornOppenheimer1927}
Max Born and Robert Oppenheimer.
\newblock Zur quantentheorie der molekeln.
\newblock \emph{Annalen der Physik}, 389\penalty0 (20):\penalty0 457--484, 1927.
\newblock \doi{10.1002/andp.19273892002}.

\bibitem[von Neumann and Wigner(1929)]{vonNeumann1929}
John von Neumann and Eugene Wigner.
\newblock Über das verhalten von eigenwerten bei adiabatischen prozessen.
\newblock \emph{Physikalische Zeitschrift}, 30:\penalty0 467--470, 1929.

\bibitem[Baer(2006)]{Baer2006}
Michael Baer.
\newblock \emph{Beyond Born-Oppenheimer: Electronic Nonadiabatic Coupling Terms and Conical Intersections}.
\newblock Wiley-Interscience, 2006.
\newblock ISBN 9780470037154.

\bibitem[Domcke et~al.(2004)Domcke, Yarkony, and Köppel]{Domcke2004}
Wolfgang Domcke, David~R. Yarkony, and Horst Köppel, editors.
\newblock \emph{Conical Intersections: Electronic Structure, Dynamics and Spectroscopy}.
\newblock World Scientific, 2004.
\newblock ISBN 9789812386106.

\bibitem[González et~al.(2012)González, Escudero, and Serrano-Andrés]{Gonzalez2012}
Leticia González, Daniel Escudero, and Luis Serrano-Andrés.
\newblock Progress and challenges in the calculation of electronic excited states.
\newblock \emph{ChemPhysChem}, 13\penalty0 (1):\penalty0 28--51, 2012.
\newblock \doi{10.1002/cphc.201100592}.

\bibitem[Yarkony(2001)]{Yarkony2001}
David~R. Yarkony.
\newblock Conical intersections: The new conventional wisdom.
\newblock \emph{Journal of Physical Chemistry A}, 105\penalty0 (26):\penalty0 6277--6293, 2001.
\newblock \doi{10.1021/jp010068n}.

\bibitem[Bravyi and Kitaev(2002)]{Bravyi2002}
Sergey Bravyi and Alexei Kitaev.
\newblock Fermionic quantum computation.
\newblock \emph{Annals of Physics}, 298\penalty0 (1):\penalty0 210--226, 2002.
\newblock \doi{10.1006/aphy.2002.6254}.

\bibitem[Seeley et~al.(2012)Seeley, Richard, and Love]{Seeley2012}
Jacob~T. Seeley, Martin~J. Richard, and Peter~J. Love.
\newblock The bravyi-kitaev transformation for quantum computation of electronic structure.
\newblock \emph{The Journal of Chemical Physics}, 137\penalty0 (22):\penalty0 224109, 2012.
\newblock \doi{10.1063/1.4768229}.

\bibitem[Belaloui et~al.(2025)Belaloui, Tounsi, Khamadja, Louamri, Benslama, Bernal~Neira, and Rouabah]{Belaloui2025}
Nacer~Eddine Belaloui, Abdellah Tounsi, Abdelmouheymen~Rabah Khamadja, Mohamed~Messaoud Louamri, Achour Benslama, David~E. Bernal~Neira, and Mohamed~Taha Rouabah.
\newblock Ground-state energy estimation on current quantum hardware through the variational quantum eigensolver: A practical study.
\newblock \emph{Journal of Chemical Theory and Computation}, 21\penalty0 (14):\penalty0 6777--6792, 2025.
\newblock \doi{10.1021/acs.jctc.4c01657}.
\newblock URL \url{https://doi.org/10.1021/acs.jctc.4c01657}.
\newblock PMID: 40581860.

\bibitem[Higgott et~al.(2019)Higgott, Wang, and Brierley]{Higgott2019}
Oscar Higgott, Daochen Wang, and Stephen Brierley.
\newblock Variational quantum computation of excited states.
\newblock \emph{Quantum}, 3:\penalty0 156, 2019.
\newblock \doi{10.22331/q-2019-07-01-156}.

\bibitem[Powell(1994)]{Powell1994}
M.~J.~D. Powell.
\newblock A direct search optimization method that models the objective and constraint functions by linear interpolation.
\newblock In S.~Gomez and Jean-Pierre Hennart, editors, \emph{Advances in Optimization and Numerical Analysis}, volume 275 of \emph{Mathematics and Its Applications}. Springer, Dordrecht, 1994.
\newblock \doi{10.1007/978-94-015-8330-5_4}.

\bibitem[Lewars(2010)]{lewars2010-basis-sets}
Errol~G Lewars.
\newblock \emph{Computational chemistry}.
\newblock Springer, Dordrecht, Netherlands, 2 edition, November 2010.

\bibitem[Watts et~al.(1989)Watts, Trucks, and Bartlett]{Watts1989}
John~D. Watts, Gary~W. Trucks, and Rodney~J. Bartlett.
\newblock The unitary coupled-cluster approach and molecular properties. applications of the ucc(4) method.
\newblock \emph{Chemical Physics Letters}, 157\penalty0 (4):\penalty0 359--366, 1989.
\newblock ISSN 0009-2614.
\newblock \doi{https://doi.org/10.1016/0009-2614(89)87262-8}.
\newblock URL \url{https://www.sciencedirect.com/science/article/pii/0009261489872628}.

\bibitem[Barkoutsos et~al.(2018)Barkoutsos, Gonthier, Sokolov, Moll, Salis, Fuhrer, Ganzhorn, Egger, Troyer, Mezzacapo, Filipp, and Tavernelli]{Barkoutsos2018}
Panagiotis~K. Barkoutsos, Jerome~F. Gonthier, Igor Sokolov, Nikolaj Moll, Gian Salis, Andreas Fuhrer, Marc Ganzhorn, Daniel~J. Egger, Matthias Troyer, Antonio Mezzacapo, Stefan Filipp, and Ivano Tavernelli.
\newblock Quantum algorithms for electronic structure calculations: Particle-hole hamiltonian and optimized wave-function expansions.
\newblock \emph{Physical Review A}, 98\penalty0 (2):\penalty0 022322, 2018.
\newblock \doi{10.1103/PhysRevA.98.022322}.

\bibitem[McClean et~al.(2016)McClean, Romero, Babbush, and Aspuru-Guzik]{McClean2016}
Jarrod~R. McClean, Jonathan Romero, Ryan Babbush, and Al{\'a}n Aspuru-Guzik.
\newblock The theory of variational hybrid quantum-classical algorithms.
\newblock \emph{New Journal of Physics}, 18\penalty0 (2):\penalty0 023023, 2016.
\newblock \doi{10.1088/1367-2630/18/2/023023}.

\bibitem[Spall(1992)]{Spall1992}
J.C. Spall.
\newblock Multivariate stochastic approximation using a simultaneous perturbation gradient approximation.
\newblock \emph{IEEE Transactions on Automatic Control}, 37\penalty0 (3):\penalty0 332--341, 1992.
\newblock \doi{10.1109/9.119632}.

\bibitem[Pellow-Jarman et~al.(2023)Pellow-Jarman, Sinayskiy, Pillay, and Petruccione]{Pellow-Jarman2023}
Aidan Pellow-Jarman, Ilya Sinayskiy, Anban Pillay, and Francesco Petruccione.
\newblock Near-term algorithms for linear systems of equations.
\newblock \emph{Quantum Information Processing}, 22:\penalty0 258, 2023.
\newblock \doi{10.1007/s11128-023-04020-2}.

\bibitem[Kraft(1988)]{Kraft1988}
Dieter Kraft.
\newblock A software package for sequential quadratic programming.
\newblock \emph{Forschungsbericht- Deutsche Forschungs- und Versuchsanstalt fur Luft- und Raumfahrt}, 1988.

\bibitem[Snyder and Mazziotti(2011)]{Snyder2011}
James W.~Jr. Snyder and David~A. Mazziotti.
\newblock Conical intersection of the ground and first excited states of water: Energies and reduced density matrices from the anti-hermitian contracted schr{\"o}dinger equation.
\newblock \emph{The Journal of Physical Chemistry A}, 115\penalty0 (48):\penalty0 14120--14126, 2011.
\newblock \doi{10.1021/jp208013m}.

\end{thebibliography}

\end{document}